\documentclass[aps,prl,showpacs,twocolumn,superscriptaddress,floatfix]{revtex4-1}

\usepackage{graphicx}
\usepackage{epstopdf}
\usepackage{color}
\usepackage[colorlinks,bookmarks=false,citecolor=blue,linkcolor=red,urlcolor=blue]{hyperref}

\usepackage{bm}

\usepackage{amsmath,amssymb}

\begin{document}

\title{Hall effect of triplons in a dimerized quantum magnet}

\author{Judit Romh\'anyi}
\affiliation{Leibniz-Institute for Solid State and Materials Research, IFW-Dresden, D-01171 Dresden, Germany }
\author{Karlo Penc}
\affiliation{Institute for Solid State Physics and Optics, Wigner Research Centre for Physics, Hungarian Academy of Sciences, H-1525 Budapest, P.O.B. 49, Hungary}
\author{R. Ganesh}
\affiliation{Leibniz-Institute for Solid State and Materials Research, IFW-Dresden, D-01171 Dresden, Germany }

\date{\today}

\begin{abstract}

SrCu$_2$(BO$_3$)$_2$ is the archetypal quantum magnet with a gapped dimer-singlet ground state and triplon excitations. It serves as an excellent realization of the Shastry Sutherland model, upto small anisotropies arising from Dzyaloshinskii Moriya (DM) interactions. We demonstrate that the DM couplings in fact give rise to topological character in the triplon band structure. 
The triplons form a new kind of Dirac cone with three bands touching at a single point, a spin-1 generalization of graphene. 
An applied magnetic field opens band gaps leaving us with topological bands with Chern numbers $\pm 2$.
SrCu$_2$(BO$_3$)$_2$ thus provides a magnetic analogue of the integer quantum Hall effect and supports topologically protected edge modes. 
At a critical value of the magnetic field set by the DM interactions, the three triplon bands touch once again in a spin-1 Dirac cone, and lose their topological character. We predict a strong thermal Hall signature in the topological regime. 
\end{abstract}


\maketitle

Topological phases of bosons have steadily gained interest, driven by the goal of realizing protected edge states that do not suffer from dissipation. 
As bosonic carriers (phonons, magnons, etc.) are electrically neutral, they are weakly interacting and show good coherent transport. 
As a first step in this direction, analogues of the integer quantum Hall effect have been proposed using photons~\citep{Raghu2008,Petrescu2012,Rechtsman2013,Hafezi2013}, magnons~\citep{Katsura2010,Shindou2013,Matsumoto2011,Ideue2012,Zhang2013}, phonons~\citep{Zhang2010,Zhang2011,Qin2012} and skyrmionic textures~\citep{Hoogdalem2013}, with the thermal Hall effect~\citep{Onose2010} as the experimental probe of choice. 
We present the first manifestation of this physics in a quantum magnet using `triplon' excitations in SrCu$_2$(BO$_3$)$_2$, the well known realization of the Shastry Sutherland model~\citep{Shastry1981,Miyahara1999}.

\begin{figure}[b]
\begin{center}
\includegraphics[width=0.9\columnwidth]{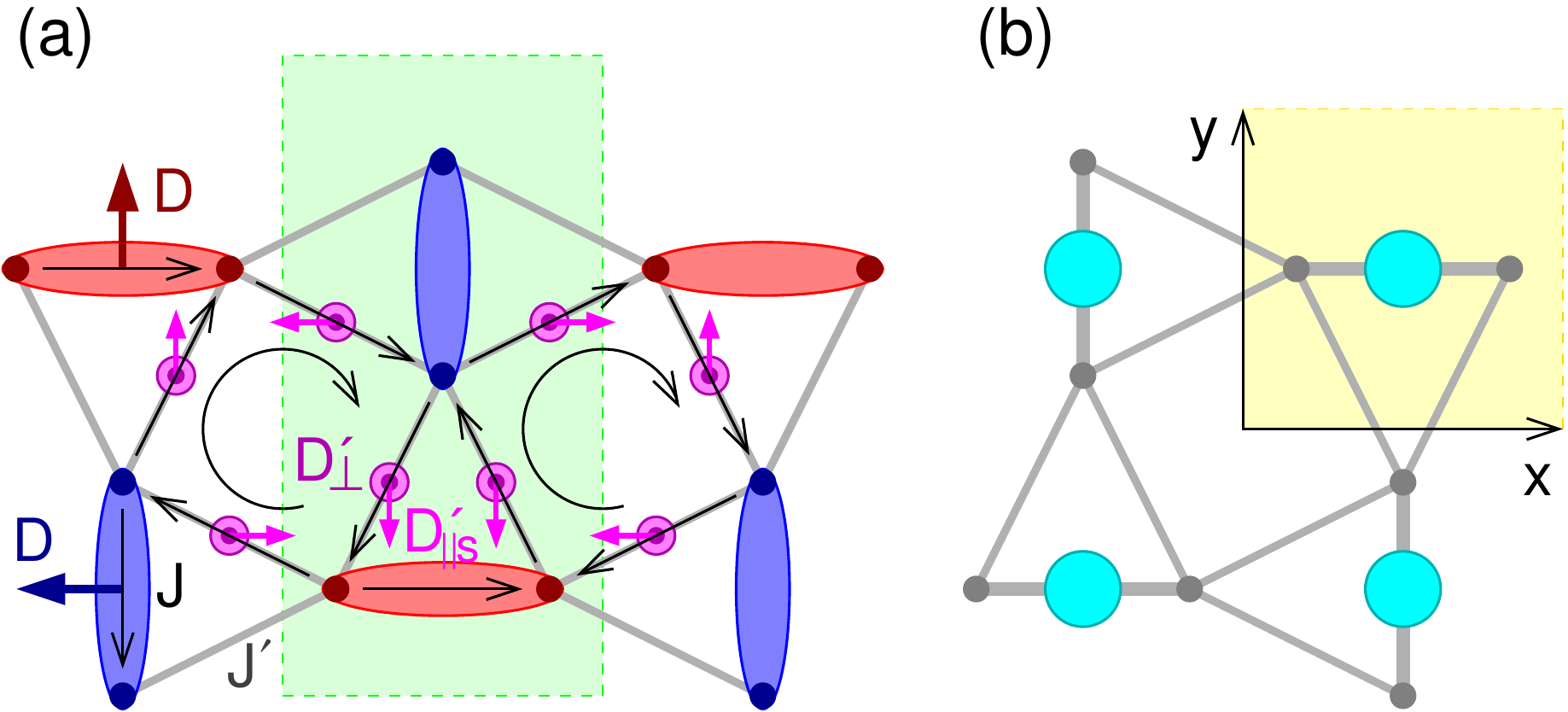}
\caption{
(a) The SrCu$_2$(BO$_3$)$_2$ lattice with Heisenberg and DM couplings. The red and blue arrows on dimers represent the intra-dimer DM vectors $\mathbf{D}$ (black arrows indicate the order of spins in the DM term). 
The inter-dimer coupling $\mathbf{D}'$ has in-plane and out-of-plane components. As we go around void squares as indicated, the out-of-plane DM component points out of the plane (purple circles). The `staggered' in-plane component is shown by magenta arrows (see text). 
The green rectangle indicates the structural unit cell.
(b) A new reduced unit cell is shown at top right, in which the two dimers are taken to be equivalent. The dimers form a square lattice as shown.
}
\label{fig:lattice}
\end{center}
\end{figure}

SrCu$_2$(BO$_3$)$_2$ is a layered material consisting of Cu $S=1/2$ moments arranged in orthogonal dimers~\citep{Smith1989,Kageyama1999}.
To a very good approximation, this arrangement conforms to the Shastry Sutherland model with spins on each dimer forming a singlet.
Low energy excitations correspond to breaking a singlet to form a triplet. Such excitations are called `triplons' and can be thought of as spin-1 \textit{bosonic} particles\cite{SachdevBhatt1990}. Indeed, triplons undergo Bose condensation in many systems\cite{Giamarchi2008}.
If SrCu$_2$(BO$_3$)$_2$ were an exact realization of the Shastry Sutherland model, the triplons would be local excitations forming a threefold-degenerate flat band~\cite{Momoi2000}.
However, electron spin resonance (ESR)~\cite{Nojiri2003}, infrared absorption (IR)~\citep{Room2004}, neutron scattering~\citep{Gaulin2004} and Raman scattering~\citep{Gozar2005} measurements show a weak dispersion that has been attributed to small Dzyaloshinskii-Moriya (DM) anisotropies~\citep{Cepas2001,Cheng2007,Romhanyi2011}. NMR measurements also support the presence of DM couplings ~\citep{Miyahara2008}.
Fig.~\ref{fig:lattice}(a) illustrates the lattice geometry and the interactions between the spins. 
 The resulting Hamiltonian is given by 
\begin{eqnarray}
\mathcal{H} &=&J\sum_{n.n.}{\bf S}_{i}{\cdot}{\bf S}_{j}
+J'\sum_{n.n.n.}{\bf S}_{i}{\cdot}{\bf S}_{j} -g_z h^z\sum_i S^z_i\nonumber\\
&& +\sum_{\text{n.n.}} \mathbf{D}_{ij}{\cdot} \left(\mathbf{S}_{i}\times\mathbf{S}_{j}\right)
+
\sum_{\text{n.n.n.}} \mathbf{D}_{ij}' {\cdot} \left(\mathbf{S}_{i}\times\mathbf{S}_{j}\right).
\label{eq.Hamiltonian}
\end{eqnarray}
We include a small magnetic field $h^z$, perpendicular to the SrCu$_2$(BO$_3$)$_2$ plane. 
The intra-dimer coupling $D$ is allowed by symmetry below a structural phase transition at $T \sim 395$~K~\citep{Smith1991,sparta2001}. 
In the inter-dimer bonds, the dominant DM component is out-of-plane. As seen in Fig.~\ref{fig:lattice}(a), the out-of-plane $D'_\bot$ couplings encode a sense of clockwise rotation; this ultimately drives a Hall effect of triplon excitations as we report below.

\begin{figure*}[t!]
\begin{center}
\includegraphics[width=0.95\textwidth]{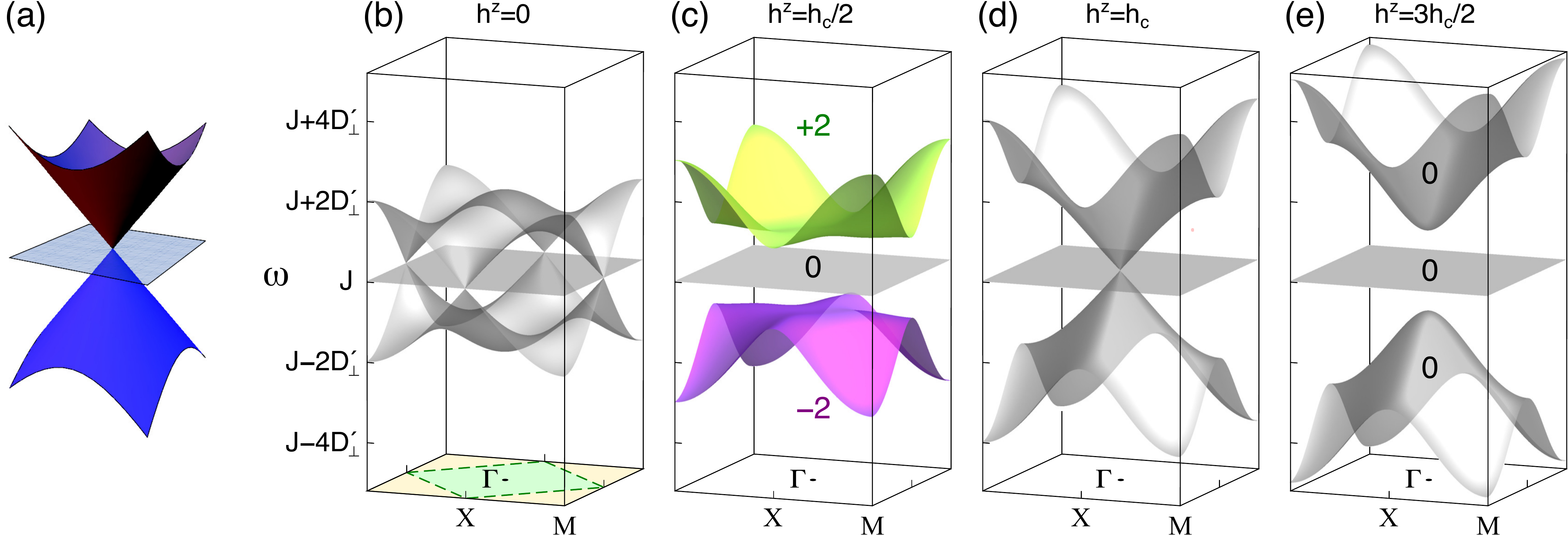}
\caption{
(a) A spin-1 Dirac cone with three bands touching. (b) Triplon dispersion for $h^z=0$. The basal plane shows the enlarged BZ corresponding to one dimer per unit cell, with $\mathbf{k}=(\pi,\pi)$ at the $M$ and $\mathbf{k}=(\pi,0)$ and $(\pi,0)$ at the $X$ points. The smaller structural BZ is shown in green. The band structure hosts spin-1 Dirac cones at the BZ edge centres X. 
(c)-(e) Evolution of triplon bands and Chern numbers upon tuning magnetic field. Bands with non-zero Chern number appear for $0<h^z<h_c$ and are shown in colour. 
At $h^z = h_c$ (d), the bands touch at a spin-1 Dirac cone at $\Gamma$. For $h^z > h_c$ as in (e), the Chern numbers remain zero.
}
\label{fig:triplon_bands}
\end{center}
\end{figure*}

\section{Methods}
\paragraph{Triplon description:}
A thorough bond operator treatment of the Hamiltonian in Eq.~(\ref{eq.Hamiltonian}) has been presented in Ref.~\citep{Romhanyi2011}. 
We present a simplified treatment suitable for SrCu$_2$(BO$_3$)$_2$ in a weak magnetic field. Previous studies have largely focussed on plateau phases at high fields(Ref. ~\citep{Matsuda2013} and references therein). In contrast, we show that the low field regime has exotic topological properties.

In a given dimer, the Hilbert space is spanned by a singlet 
$\vert s \rangle  = (\vert\!\uparrow \downarrow \rangle - \vert\!\downarrow \uparrow \rangle)/\sqrt{2}$ 
and three triplets: 
$\vert t_{x}\rangle = i(\vert\!\uparrow \uparrow \rangle - \vert \!\downarrow \downarrow \rangle)/\sqrt{2} $, 
$\vert t_{y}\rangle = (\vert\!\uparrow \uparrow \rangle + \vert \!\downarrow \downarrow \rangle)/\sqrt{2}$ 
and 
$\vert t_z\rangle = -i(\vert\! \uparrow \downarrow \rangle + \vert\! \downarrow \uparrow \rangle)/\sqrt{2}$.
In the pure Shastry-Sutherland model, the ground state is a direct product of singlets $|s\rangle$ over the dimers as long as $J' \lesssim 0.675 J$~\citep{Miyahara1999,Koga2000,Corboz2013}. 
In SrCu$_2$(BO$_3$)$_2$, as the DM anisotropies are small compared to $J$, we assume that the ground state remains a product wavefunction.
Minimizing the overall energy, we find that the ground state has the wavefunction
$\vert \tilde{s} \rangle_{\sf h}
\sim \vert s \rangle_{\sf h}- \alpha \vert t_y \rangle_{\sf h}$ and 
$\vert \tilde{s} \rangle_{\sf v} 
\sim \vert s \rangle_{\sf v} + \alpha \vert t_x \rangle_{\sf v}$
on horizontal and vertical dimers, respectively; the direction of $\mathbf{D}$ on each dimer determines whether $\vert t_y \rangle$ or $\vert t_x \rangle$ is admixed.
The triplet admixture is proportional to the intra-dimer DM coupling $D$ with $\alpha \approx D/2J\ll 1$. Here, as in the rest of this article, we only retain terms up to linear order in $D,D'$, and $h^z$ which are small compared to the $J'$s. 

On each dimer, we choose a new Hilbert space by rotating   
 $\mathbf{w}_i=(|s\rangle,|t_x\rangle,|t_y\rangle,|t_z\rangle)_i$ to $ \mathbf{\tilde w}_i = W_{{\sf h}/{\sf v}} \cdot \mathbf{w}_i$ using 
 \begin{align}
W_{\sf h}
=
\left(
\begin{array}{cccc}
 1 & 0 & -\alpha  & 0 \\
 0  &  1 & 0 & 0 \\
 \alpha &  0  &1  & 0 \\
 0 & 0  &  0 & 1 
\end{array}
\right)
\;\text{and}\;
W_{\sf v}
=
\left(
\begin{array}{cccc}
 1 & \alpha  & 0 & 0 \\
 i\alpha& -i & 0   & 0 \\
 0 &  0 & -i & 0 \\
 0 & 0  &  0 & i 
\end{array}
\right).
\end{align}
on horizontal and vertical dimers respectively. In the ground state, each dimer is in the $\vert \tilde{s}\rangle$ state given by the first row in the corresponding W matrix. We have three local excitations given by the mutually orthogonal `triplon' states $\vert \tilde{t}_x \rangle$ , $\vert \tilde{t}_y \rangle$ and $\vert \tilde{t}_z \rangle$.

At low magnetic fields, the low-energy excitations are spanned by single-triplon states with their dynamics captured by hopping processes of the form $_i\langle \tilde{t}_{\alpha}\vert \mathcal{H}\vert \tilde{t}_\beta\rangle_{j}$. Introducing a bosonic representation for triplons, we obtain a Hamiltonian with purely hopping-like terms. 
By defining $W_{\sf v}$ as above with complex entries, the Hamiltonian takes on a convenient form, viz., the two dimers in the unit cell become equivalent (see Supplementary Note 2 for details). We may henceforth drop ${\sf v}/{\sf h}$ indices and work with the reduced unit cell in Fig.~\ref{fig:lattice}(b). In momentum space, the Brillouin zone (BZ) is enlarged as shown in Fig.~\ref{fig:triplon_bands}(b).

For a more complete treatment, we may include pairing-like terms ($\tilde{t}_{i,\alpha}^\dagger \tilde{t}_{j,\beta}^\dagger$) within a bond operator formalism as in Ref.~\cite{Romhanyi2011}. We ignore such terms as they do not change the triplon energies to linear order in D, D' and $h^z$; we have checked that their inclusion does not alter the results presented here.  

\section{Results}
\paragraph{Spin-1 Dirac cone physics:}

The triplon Hamiltonian in momentum space is given by
\begin{eqnarray}
\label{eq:Hamilton}
\mathcal{H}=\sum_{\mathbf{k}}\sum_{\mu,\nu=x,y,z}
\tilde{t}^{\dagger}_{\mu,\mathbf{k}}
M^{\phantom{\dagger}}_{\mu\nu}({\mathbf{k}})
\tilde{t}^{\phantom{\dagger}}_{\nu,\mathbf{k}},
\end{eqnarray}
where the Hamiltonian matrix is given by 
\begin{equation}
\label{eq:M}
M({\mathbf{k}}) \!= \!
\left(\!\!\!
\begin{array}{ccc}
 J & i h^z g_z + 2i D'_\bot \gamma_{3}  & \tilde D_{\|}\gamma_{2} \\
-i h^z  g_z - 2i D'_\bot \gamma_{3} & J  & -\tilde D_{\|}\gamma_{1} \\
\tilde D_{\|}\gamma_{2}  & -\tilde D_{\|}\gamma_{1} & J
\end{array}\!\!\!\right)\!\!,
\end{equation}
with
$\gamma_{1} ({\mathbf{k}}) = \sin k_x$, $\gamma_{2} ({\mathbf{k}}) = \sin k_y$, and $\gamma_{3} ({\mathbf{k}}) =\frac{1}{2} (\cos k_x+\cos k_y)$ (see Supplementary Note 2 for details). Only two components of the inter-dimer DM coupling enter the Hamiltonian, viz., the out-of-plane component $D'_\bot$ and the `staggered' component shown in Fig. \ref{fig:lattice}a. A third non-staggered component is allowed by symmetry, but does not appear at this level (see Supplementary Note 1).
Intradimer $D$ and in-plane interdimer $D'_{\|,s}$ act in consonance so that only the linear combination 
$\tilde D_{\|}=D'_{\|,s}-\frac{D J'}{2J}$ appears in the Hamiltonian similar to the analysis in Ref.~\cite{Cheng2007}.
In the following analysis, we use the values $J=722$~GHz, 
$J' = 468$~GHz, 
$|\tilde D_{\|}| = 20 $~GHz , $D_\bot'  = -21$~GHz and $g_z=2.28$ in the $M({\mathbf{k}})$ matrix, which reproduce the ESR data in Ref.~[\onlinecite{Nojiri2003}]. The parameter $J$ is not the microscopic exchange strength, but rather the measured spin gap which determines the effective coupling in the presence of quantum fluctuations. 

The $M({\mathbf{k}})$ matrix is of the form 
\begin{equation}
M({\mathbf{k}}) =  J \mathbf{1} + \mathbf{d}({\mathbf{k}})\cdot \mathbf{L} \;,
\label{Eq.Hform}
\end{equation}
where $\mathbf{1}$ is the $3\times 3$ identity matrix and
\begin{eqnarray}
\mathbf{L} = \left[ 
\left(\begin{array}{ccc}
0 & 0 & 0\\
0 & 0 & -1\\
0 & -1 & 0
\end{array}\right),
\left(\begin{array}{ccc}
0 & 0 & 1\\
0 & 0 & 0\\
1 & 0 & 0
\end{array}\right),
\left(\begin{array}{ccc}
0 & -i & 0\\
i & 0 & 0\\
0 & 0 & 0
\end{array}\right)\right]
\end{eqnarray}
is a vector of $3\times 3$ matrices satisfying the $[L^\xi,L^\eta] = i \varepsilon_{\xi\eta\zeta} L^\zeta$ SU(2) algebra. 
Thus, in momentum space, the triplons behave as (pseudo)spin-1 objects coupled to a pseudomagnetic field 
\begin{equation}
 \mathbf{d}({\mathbf{k}})=\left[\tilde D_{\|}\gamma_{1}({\mathbf{k}}),\tilde D_{\|}\gamma_{2}({\mathbf{k}}), -h^z g_z - 2  D'_\bot \gamma_{3}({\mathbf{k}}) \right] \;. 
\end{equation}

We now draw an analogy with the usual two-band physics wherein the $2\times 2$ Hamiltonian takes the same form as Eq.~(\ref{Eq.Hform}) but with spin-1/2 Pauli matrices instead of spin-1 $\mathbf{L}$ matrices. 
There, we obtain two bands corresponding to eigenvalues $J\pm d({\mathbf{k}})/2$ (we denote $d({\mathbf{k}})=|\mathbf{d}({\mathbf{k}})|$). 
If $\mathbf{d}({\mathbf{k}})$ is non-zero throughout the BZ, we obtain two well separated bands whose Chern numbers are $\pm N_s$, where $N_s$ is the number of skyrmions in the $\mathbf{d}({\mathbf{k}})$ field over the BZ\cite{Bernevig}. 
The $\mathbf{d}({\mathbf{k}})$ field contains all information about the band structure; its skyrmion count determines the topological character of bands.
We emphasize here that topological properties will not change with small corrections to the Hamiltonian such as next-nearest neighbour hopping (see Supplementary Note 5).

Likewise, in our spin-1 realization, we read off the eigenvalues as $\{J+d({\mathbf{k}}), J, J-d({\mathbf{k}})\}$. 
Note that the band in the middle is always flat with energy $J$, irrespective of the value of $\mathbf{d}({\mathbf{k}})$. 
If the pseudomagnetic field $\mathbf{d}({\mathbf{k}})$ vanishes at some ${\mathbf{k}}$, all three bands touch in a `spin-1 Dirac cone', resembling graphene but with an additional flat band passing through the band touching point.
If $\mathbf{d}({\mathbf{k}})$ is non-zero throughout the BZ, the spectrum consists of
three well-separated triplon bands 
with well-defined Chern numbers 
$\{-2 N_s, 0, +2N_s\}$, where $N_s$ is again the skyrmion number. More generally, for the arbitrary spin-$S$ generalization of Eq.~\ref{Eq.Hform}, we have $(2S+1)$ bands with Chern numbers $\{-2S N_s, -2(S-1)N_s, \cdots, 2(S-1), 2SN_s\}$ (see Supplementary Note 3). 

\paragraph{Magnetic field tuned topological transitions:}
 
The magnetic field $h^z$ provides a handle to tune topological transitions in SrCu$_2$(BO$_3$)$_2$, as shown in 
Fig.~\ref{fig:triplon_bands}. With small magnetic fields, even though the ground state remains a product of dimer singlets, the band structure of excitations shows topological transitions.
When $h^z=0$, the three bands touch at 
the edge centres of the BZ (corresponding to corners in the structural BZ). 
A small applied field opens a non-trivial band gap, 
allowing for three well-separated bands with Chern numbers $\{-2,0,+2\}$ or $\{+2,0,-2\}$, depending on the sign of $h^z$. 
When the field reaches a critical strength $h_c = 2|D'_\bot|/g_z$, the three bands touch at the  $\Gamma$ point. Indeed, this band touching has already been seen in ESR ~\citep{Nojiri2003} and infrared absorption ~\citep{Room2004} spectra at $h^z\approx 1.4$~T; however, its significance as a spin-1 Dirac point was not appreciated.
As $h^z$ is increased further, a trivial band gap opens with all three Chern numbers being zero. 

The topology of triplon bands can be understood in terms of the $\mathbf{d}({\mathbf{k}})$ field. To every point in the 2D BZ (an $\mathbb{S}^1 \times \mathbb{S}^1$ torus), we assign the 3D vector $\mathbf{d}({\mathbf{k}})$: this gives us a closed 2D surface embedded in 3 dimensions. If the bands are to remain well-separated, the surface cannot touch the origin, {\it i.e.} $\mathbf{d}({\mathbf{k}})\neq 0$ anywhere in the BZ. 
The origin is thus special and acts as a monopole for Berry phase. 
The topology of the band structure reduces to whether or not the 2D surface encloses the origin; if it does, how many times does it wrap around the origin? This defines a skyrmion number $N_s \in \mathbb{Z}$, that is related to the Chern number. 

\begin{figure}[h!]
\begin{center}
\includegraphics[width=0.95\columnwidth]{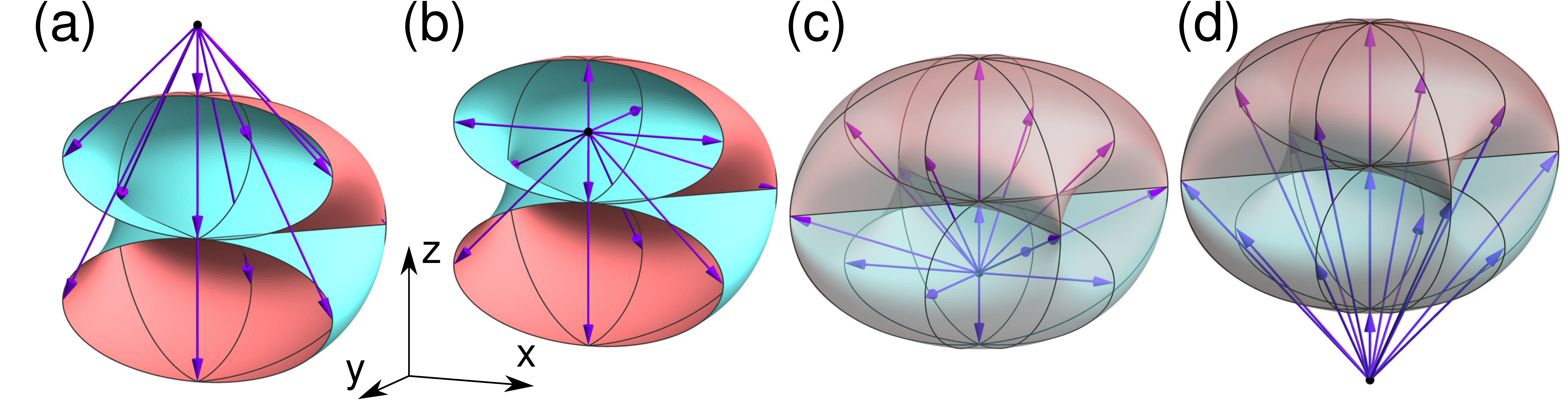}
\caption{ Topological 
2D surface obtained from mapping each point in the BZ to a 3D $\mathbf{d}(\mathbf{k})$ vector, for
$h^z = -3h_c/2$, $-h_c/2$, $h_c/2$, and $3h_c/2$ (from left to right).
The arrows in the figure emanate from the origin which acts as a monopole of Berry flux. 
}
\label{fig:skyrmion}
\end{center}
\end{figure}

To see the role of $h^z$, we note that it enters solely as an additive contribution in the $z$-component of $\mathbf{d}({\mathbf{k}})$. 
As shown in Fig.~\ref{fig:skyrmion},
the BZ maps to a closed surface of width $2\tilde D_{\|}$ and height $4D'_\bot$, which is composed of an upper and a lower chamber. The chambers are disconnected, but touch along line nodes. The surface is orientable: the outer surface of the lower chamber smoothly connects to the inner surface of the upper chamber and vice versa.
When $\vert h^z \vert > h_c$, neither chamber encloses the origin; we have $N_s = 0$ with all Chern numbers zero [Figs.~\ref{fig:skyrmion}(a) and (d)]. When $-h_c < h^z < 0$, the origin lies inside the upper chamber [Fig.~\ref{fig:skyrmion}(b)], the net Berry flux is positive and Chern numbers are $\{+2,0,-2\}$. When $0 < h^z < h_c$, the origin lies inside the lower chamber [Fig.~\ref{fig:skyrmion}(c)], the Berry flux is negative and Chern numbers are $\{-2,0,+2\}$.

The key ingredient that gives rise to topological properties is the DM interaction that originates from relativistic spin-orbit coupling.
The critical magnetic field $h_c$ is proportional to the coupling $D'_\bot$. 
The intra-dimer DM coupling $D$ also plays a role: we do not find any Chern bands upon setting $D=0$, as is appropriate for $T > 395$~K, above a structural transition in SrCu$_2$(BO$_3$)$_2$.

\begin{figure}[h!]
\begin{center}
\includegraphics[width=0.95\columnwidth]{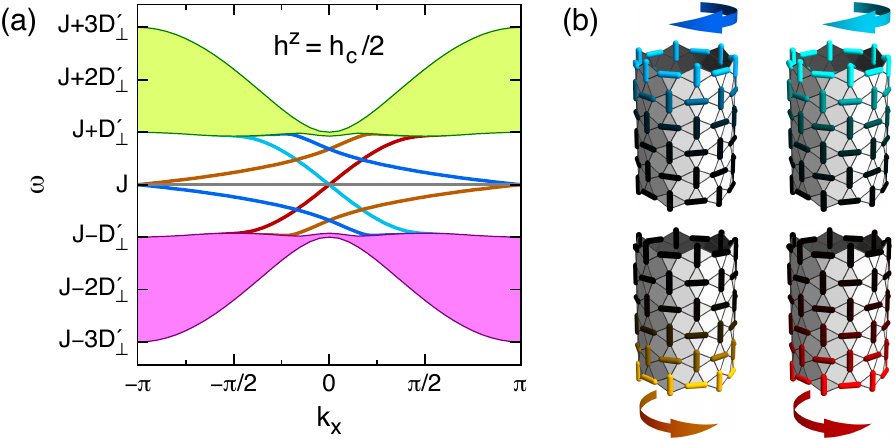}
\caption{ 
(a) Band structure of SrCu$_2$(BO$_3$)$_2$ on a cylindrical strip similar to the one shown in (b) upon taking the width to be very large. We recover the bulk states of Fig.~\ref{fig:triplon_bands}(c). In addition, four edge states appear, connecting the Chern bands. 
(b) Wavefunctions of the four edge states for an arbitrary $k_x$ on a strip of width $W=8$ dimers. The color of the dimer bond represents the triplon weight with black corresponding to zero. Right-moving edge states are localized on the bottom edge while left-movers are localized on the top edge.}
\label{fig:edges}
\end{center}
\end{figure}

\paragraph{Edge states:}
The topological character of bands is revealed when edges are introduced. 
For $0<h^z<h_c$ (and for $-h_c<h^z<0$), edge states connecting the Chern bands appear within the bulk band gap, as shown in 
Fig.~\ref{fig:edges}(a) for a strip geometry. 
Apart from recovering the bulk bands, we clearly see \textit{four }edge states consistent with bulk boundary correspondence~\citep{Hatsugai1993} for Chern numbers $\pm 2$. 
The edge states constitute two `right-movers' and two `left-movers' (with group velocity pointing right/left), localized on the opposite edges of the strip. 
The wave functions of the edge states decay exponentially into the bulk, as shown in [Fig.~\ref{fig:edges}(b)]

\begin{figure}
\begin{center}
\includegraphics[width=0.95\columnwidth]{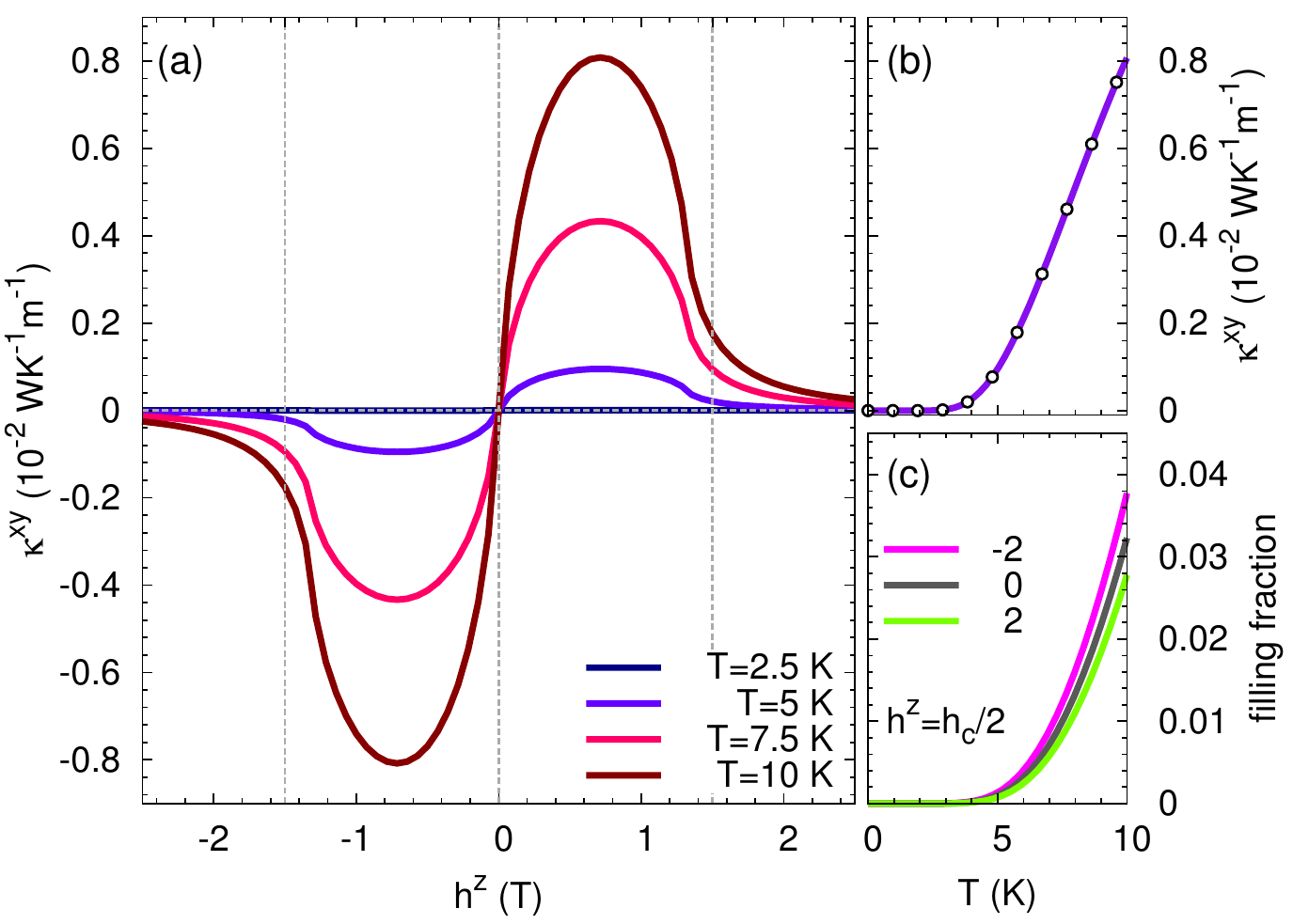}
\caption{
(a) Thermal Hall conductivity vs. external magnetic field at different temperatures. The critical fields $h^z = \pm h_c$ are shown as vertical dashed lines. 
(b) Thermal Hall signal at $h^z = h_c/2$ vs. temperature. 
The circles are direct evaluation of the formula, the line is an approximation applicable to SrCu$_2$(BO$_3$)$_2$ (see Supplementary Note 4 for derivation taking bandwidths to be much smaller than the gap). 
(c) The filling fraction (boson occupation number) for the three bands, indexed by Chern numbers. It grows much slower with temperature than $\kappa^{xy}$. 
}
\label{fig:Hall}
\end{center}
\end{figure}

\paragraph{Thermal Hall effect:}
Chern bands in electronic systems can be easily probed by doping the system so that the Fermi level lies in the band gap. 
This gives a transverse electrical conductivity quantized to integer values. 
In bosonic systems where this is not possible, the thermal Hall effect provides an alternative. 
Semi--classical analysis shows that a wave packet in a Chern band undergoes rotational motion~\citep{Sundaram1999,Niu2010}. 
To exploit this, a temperature gradient is used to populate the band differently at the system's edges. 
The rotational motion of the triplons is then unbalanced, leading to a transverse triplon current. 
As triplons carry energy, this leads to a measurable transverse thermal current.

An expression for thermal Hall conductivity was derived using the Kubo formula in Ref.~\citep{Katsura2010}. 
Subsequently, Matsumoto et al.~\citep{Matsumoto2011} showed that there is an extra contribution from the orbital motion of excitations. 
Fig.~\ref{fig:Hall}(a) shows the thermal Hall conductivity as a function of external magnetic field calculated using the expression in Ref.~\citep{Matsumoto2011}. 
SrCu$_2$(BO$_3$)$_2$ is quasi-two-dimensional and the Hall response in each layer is in the same direction. Therefore, we add the contribution from each layer to get $\kappa^{xy}$ for a three dimensional sample. 
As the magnetic field is tuned away from $h^z=0$, a non-zero Hall signal develops with the sign of $\kappa^{xy}$ depending on the direction of magnetic field.
When the critical magnetic field strength $h_c$ is reached, the topological nature of triplon bands is lost and the Hall signal is diminished. 
Fig.~\ref{fig:Hall}(b) shows the peak thermal Hall conductivity increasing monotonically with background temperature. 
Our calculation assumes that the temperature is low enough that the triplon bands are weakly populated, allowing us to neglect triplon-triplon interactions. 
We expect this assumption to hold atleast until $\sim 5$~K where the filling of bosons is $\sim0.2\%$. Neutron scattering data shows that the intensity of the single triplet excitations is essentially unchanged up to 5 K showing no damping.\cite{Gaulin2004}.

\section{Discussion}
We have demonstrated that SrCu$_2$(BO$_3$)$_2$ hosts a Hall effect of triplons. A small external magnetic field of the order of a few Tesla suffices to tune topological transitions in the band structure. 
The triplons form novel spin-1 Dirac cones with threefold band touching. 
Such a feature has been seen in various contexts~\cite{Apaja2010,Huang2011,Asano2011,Dora2011,Yamashita2013}. 
Our study elucidates its implications for band structure topology; the spin-1 structure naturally gives Chern numbers $\pm 2$ instead of the more common $\pm 1$. 
Similar topological phases could exist in dimer compounds such as Rb$_2$Cu$_3$SnF$_{12}$~\citep{Matan2010,Kyusung2012} with non-zero DM couplings, and possibly in ZnCu$_3$(OH)$_6$Cl$_2$ (Herbertsmithite)~\citep{Tover2009}. 

We predict a thermal Hall signature in SrCu$_2$(BO$_3$)$_2$ that can be verified by transport measurements. 
We also suggest neutron scattering experiments to study the evolution of band structure in low magnetic fields ($\lesssim$ 2T).
Such measurements can see the spin-1 Dirac cone features at $h^z=0$ and $h^z = h_c$. 
It may even be possible to directly probe the edge states using precise low-angle scattering measurements. 

\section{Acknowledgements}
We thank R. Shankar (Chennai), A. Paramekanti and M. Daghofer for useful discussions.
This work was supported by Hungarian OTKA Grant No. 106047.

%
%


\clearpage

\renewcommand{\theequation}{S.\arabic{equation}}
\renewcommand{\thefigure}{S.\arabic{figure}}
\renewcommand{\thetable}{S.\Roman{table}}
\renewcommand{\thesection}{Supplementary Note \arabic{section}}
\renewcommand{\thesubsection}{Supplementary Note \arabic{subsection}}

\setcounter{equation}{0}
\setcounter{figure}{0}
\setcounter{table}{0}

\subsection{Supplementary Notes\\ \vspace{1ex}1. Dzyaloshinsky-Moriya interactions allowed by symmetry}
\label{sec:DMinter}

\begin{figure}[h!bt]
\begin{center}
\includegraphics[width=0.7\columnwidth]{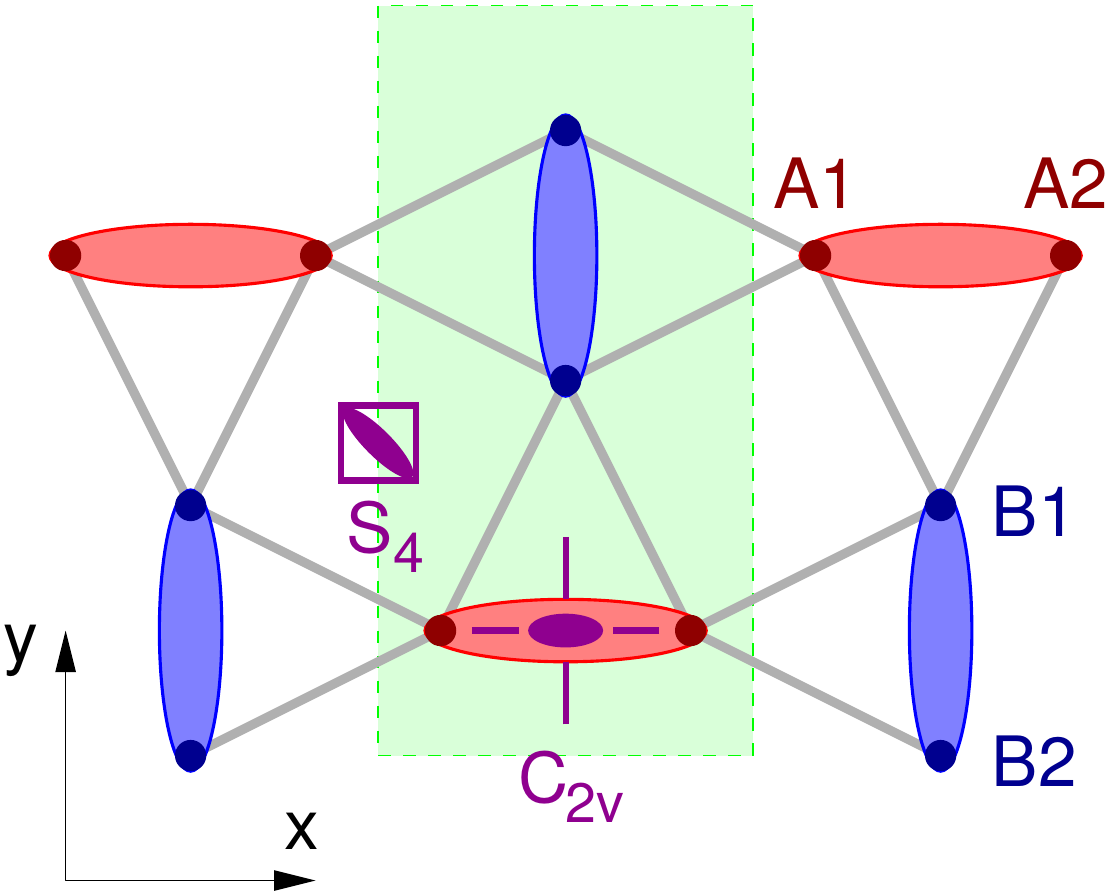}
\caption{Site symmetries in the buckled CuBO$_3$ layer: $\mathcal{S}_4$ in the centre of four dimer and $\mathcal{C}_{2v}$ in the centre of the dimers. Copper ions below and above the layer are indicated by red and blue circles, respectively. The green rectangle indicates the unit cell.
}
\label{fig:struct}
\end{center}
\end{figure}

The Dzyaloshinskii-Moriya couplings in SrCu$_2$(BO$_3$)$_2$ have been discussed by several authors\cite{Choi2003,Kodama2005,Cheng2007}. Here, we present a systematic symmetry-based derivation of the correct DM vectors. 
SrCu$_2$(BO$_3$)$_2$ undergoes a structural transition at $T_{s}=395$ K, when the dimers shift in opposite directions perpendicular to the plane. Below $T_s$, with the loss of inversion symmetry, the space group is I$\bar42$m. 
The unit cell consists of two orthogonal dimers: dimer $A$ which is parallel to the $x$-axis, and dimer $B$ parallel to the $y$-axis, as indicated in Fig. \ref{fig:struct}.
The symmetry group of the unit cell for the low temperature structure is isomorphic to $\mathcal{D}_{2d}$, consisting of 8 symmetry elements: $E$, $\sigma_{xz}$, $\sigma_{yz}$, $C_2$, $S_4$, $S_4^3$, $\sigma_{xz} S_4$ and $\sigma_{yz} S_4$. The rotation axis of $\mathcal{S}_4$ is pinned to the center of four sites, while the mirror planes, together with the $C_2(z)$ rotation, constitute the $\mathcal{C}_{2v}$ on the centres of dimers, as illustrated in Fig. \ref{fig:struct}. The effects of these symmetry elements on the sites and on the spin components are given in Table \ref{tab:group_action}. By examining the symmetry operations, we can construct invariant combinations of spin operators that are allowed in the Hamiltonian.

\begin{table}[bt]
\begin{center}
\begin{tabular}{crrrrrrr}
$E$ & $\sigma_{xz}$ & $\sigma_{yz}$ & $C_2$ & $S_4$ & $S_4^3$ & $\sigma_{xz} S_4$ & $\sigma_{yz} S_4$ \\
\hline 
\hline 
$S^x$ &$-S^x$ & $S^x$ &$-S^x$ &$-S^y$ & $S^y$ &$-S^y$ & $S^y$ \\
$S^y$ & $S^y$ &$-S^y$ &$-S^y$ & $S^x$ &$-S^x$ &$-S^x$ & $S^x$ \\
$S^z$ &$-S^z$ &$-S^z$ & $S^z$ & $S^z$ & $S^z$ &$-S^z$ &$-S^z$ \\
\hline
A1 & A1 & A2 & A2 & B2 & B1 & B1 & B2 \\
A2 & A2 & A1 & A1 & B1 & B2 & B2 & B1 \\
B1 & B2 & B1 & B2 & A1 & A2 & A1 & A2 \\
B2 & B1 & B2 & B1 & A2 & A1 & A2 & A1 \\
\hline 
\end{tabular}
\caption{\label{tab:group_action} The transformation of spin components and sites under the symmetry transformations of the point group of the unit cell in the low symmetry case. }
\end{center}
\end{table}

The anisotropies arising from intradimer (${\bf D}$) and interdimer (${\bf D'}$) Dzyaloshinsky-Moriya  interactions take the form:
\begin{equation}
\mathcal{H_{\rm DM}}= 
 \sum_{nn} \mathbf{D}_{ij}\cdot\left(\mathbf{S}_{i}\times\mathbf{S}_{j}\right)
 +\sum_{nnn} \mathbf{D'}_{ij}\cdot\left(\mathbf{S}_{i}\times\mathbf{S}_{j}\right)
\label{eq:HDM}
\end{equation}
The symmetry properties of the lattice determine the allowed components of the vectors ${\bf D}$ and ${\bf D'}$.

The intradimer interaction on the bond type $A$ has the form of  $\mathbf{D}_A\left(\mathbf{S}_{A1}\times\mathbf{S}_{A2}\right)$ that can be written as a determinant:
\begin{equation}
\mathbf{D}_A\cdot\left(\mathbf{S}_{A1}\times\mathbf{S}_{A2}\right) = 
\left| \begin{array}{ccc}
D^x_A & D^y_A & D^z_A\\
S^x_A & S^y_A & S^z_A\\
S^x_A & S^y_A & S^z_A
\end{array}\right| \;.
\end{equation}
This determinant must be invariant under all the symmetry elements of $\mathcal{D}_{2d}$. For example, upon applying $C_2$,
\begin{align}
C_2 \left| \begin{array}{ccc}
D^x_A & D^y_A & D^z_A\\
S^x_{A1} & S^y_{A1} & S^z_{A1}\\
S^x_{A2} & S^y_{A2} & S^z_{A2}
\end{array}\right| & =
\left| \begin{array}{ccc}
D^x_A & D^y_A & D^z_A\\
-S^x_{A2} & -S^y_{A2} & S^z_{A2}\\
-S^x_{A1} & -S^y_{A1} & S^z_{A1}
\end{array}\right| 
\nonumber\\&
 =
\left| \begin{array}{ccc}
D^x_A & D^y_A & -D^z_A\\
S^x_{A1} & S^y_{A1} & S^z_{A1}\\
S^x_{A2} & S^y_{A2} & S^z_{A2}
\end{array}\right| \;.\nonumber\\
\end{align}
The original determinant and the one after applying $C_2$ must be equal, therefore it follows that $D^z_A=0$. Similarly, applying $\sigma_{xz}$ results in $D^x_A=0$. Application of $\sigma_{yz}$ does not give a new condition. However, $S_4$, a rotation followed by inversion (in accordance with Table~\ref{tab:group_action}), gives:
\begin{align}
S_4\left| \begin{array}{ccc}
D^x_A & D^y_A & D^z_A\\
S^x_{A1} & S^y_{A1} & S^z_{A1}\\
S^x_{A2} & S^y_{A2} & S^z_{A2}
\end{array}\right| &=
\left| \begin{array}{ccc}
D^x_A & D^y_A & D^z_A\\
-S^y_{B2} & S^x_{B2} & S^z_{B2}\\
-S^y_{B1} & S^x_{B1} & S^z_{B1}
\end{array}\right|
\nonumber\\ & =
\left| \begin{array}{ccc}
 -D^y_A & D^x_A & -D^z_A\\
 S^x_{B1} & S^y_{B1} &S^z_{B1}\\
S^x_{B2}  & S^y_{B2} &  S^z_{B2}
\end{array}\right|\nonumber \;.
\end{align}
This provides $D^y_A=-D^x_B$, $D^x_A=D^y_B$ and $D^z_A=D^z_B$. Since we have already established that  $D^x_A=0$ and $D^z_A=0$, we obtain $D^y_B=0$ and $D^z_B=0$. The remaining transformations do not give any new conditions, thus we conclude that the form of the intradimer Dzyaloshinsky-Moriya interaction in the unit cell is:
\begin{equation}
\mathcal{H^{{\rm intra}}_{\rm DM}}= 
D\left(\mathbf{S}_{A1}\times\mathbf{S}_{A2}\right)_y-D\left(\mathbf{S}_{B1}\times\mathbf{S}_{B2}\right)_x \;.
\end{equation}
\begin{figure}[h!tb]
\begin{center}
\includegraphics[width=0.7\columnwidth]{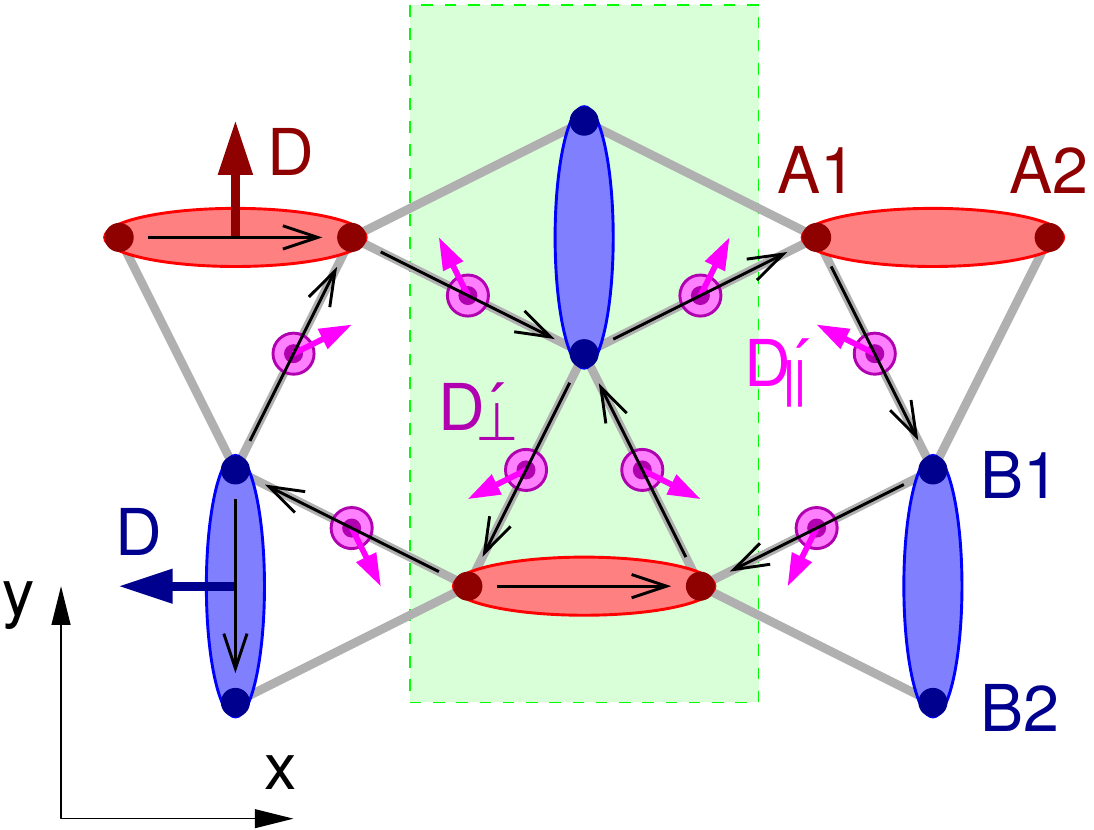}
\caption{Symmetry allowed components of the DM vectors for $T<T_s$. The intradimer DM vectors are ${\bf D}_{A}=(0,D,0)$ and ${\bf D}_B=(-D,0,0)$. The interdimer DM vector can have arbitrary components. Once we specify its direction on a given bond, though, the DM interactions on the remaining bonds in the unit cell follow from the symmetry properties.
}
\label{fig:DM_D2d}
\end{center}
\end{figure}

The symmetries above do not constrain the inter-dimer DM vector, it can point in an arbitrary direction: ${\mathbf D}'=(-D'_{\parallel ns},D'_{\parallel s},D'_{\bot})$, here multiplying $\left(\mathbf{S}_{A1}\times\mathbf{S}_{B1}\right)$. However, fixing this vector on one bond, the directions of the remaining DM vectors are determined by symmetry, as illustrated in Fig. \ref{fig:DM_D2d}. Following the notation of Ref.~[\onlinecite{Cheng2007}], we denote one of the in-plane components as `staggered' ($D'_{\parallel s}$) and the other as `non-staggered' ($D'_{\parallel ns}$). When we go around a triangle of nearest neighbour bonds (one intra-dimer bond and two inter-dimer bonds), the staggered component is the same on the two inter-dimer bonds whereas the non-staggered component switches sign. Only the staggered component enters the triplon Hamiltonian defined in Eq. 4 of the main article. 
We note that our DM vectors are slightly different from those obtained in Ref. \onlinecite{Cheng2007}, however they are fully consistent with lattice symmetries as shown above.

\subsection{\\ \vspace{1ex}2. Construction of the $3\times3$ matrix for the dispersion of the triplons }

The Hamiltonian matrix in the $\mathbf{k}$ space is na\"ively a $6\times6$ matrix, as we have three triplons each on the A and B dimers in the unit cell. However, when the magnetic field is along the $z$ direction, we can construct a unitary transformation that makes the hopping Hamiltonian translationally invariant with a single dimer per unit cell, leading to a much simpler $3\times3$ Hamiltonian matrix.
 We show this construction below. 

As first step, we construct the eigenstates of the single-dimer problem including the intradimer DM interaction. This is achieved by the unitary transformation
\begin{equation}
\left(\begin{array}{c}
s^{\prime}_{\mathbf{r}} \\
t^{\prime\phantom{\dagger}}_{x,\mathbf{r}}\\ 
t^{\prime\phantom{\dagger}}_{y,\mathbf{r}} \\
t^{\prime\phantom{\dagger}}_{z,\mathbf{r}} 
\end{array}\right)
= \mathbf{V}(\mathbf{r})
\cdot
\left(\begin{array}{c}
s^{\phantom{\dagger}}_{\mathbf{r}} \\
t^{\phantom{\dagger}}_{x,\mathbf{r}}\\ 
t^{\phantom{\dagger}}_{y,\mathbf{r}} \\
t^{\phantom{\dagger}}_{z,\mathbf{r}} 
\end{array}\right),
\end{equation}
Keeping terms up to linear order in $D/J$, the $\mathbf{V}(\mathbf{r})$ matrices are
\begin{align}
\mathbf{V}_A =& 
\left(\begin{array}{cccc}
 1 & 0 & - \frac{D}{2J} & 0  \\
  0 & 1 & 0 & 0\\
 + \frac{D}{2J} & 0 & 1 & 0\\
0 & 0 & 0 & 1\\
\end{array}\right), 
\\
\mathbf{V}_B = & 
\left(\begin{array}{cccc}
 1 &  \frac{D}{2J} & 0 & 0  \\
 -\frac{D}{2J} & 1 & 0 & 0\\
 0 & 0 & 1 & 0\\
0 & 0 & 0 & 1\\
\end{array}\right),
\end{align}
on the A and B bonds, respectively.

In real space, the Hamiltonian for the triplons is $\mathcal{H} = \mathcal{H}_{\text{site}}+\mathcal{H}_{\text{hop}}$, where
\begin{equation}
\mathcal{H}_{\text{site}} = 
\sum_{\mathbf{r}} 
\left[
J
\sum_{\mu=x,y,z}
t^{\prime\dagger}_{\mu,\mathbf{r}} 
t^{\prime\phantom{\dagger}}_{\mu,\mathbf{r}} 
+ i h^z \left( t^{\prime\dagger}_{x,\mathbf{r}} t^{\prime\phantom{\dagger}}_{y,\mathbf{r}} - t^{\prime\dagger}_{y,\mathbf{r}} t^{\prime\phantom{\dagger}}_{x,\mathbf{r}} \right) \right]
\end{equation}
is the single-dimer term
and 
\begin{align}
\label{eq:Hamilton}
\mathcal{H}_{\text{hop}}=&
\sum_{\mathbf{r}\in\Lambda_A}
\sum_{{\bm\delta}}
\sum_{\mu,\nu=x,y,z}
t^{\prime\dagger}_{\mu,\mathbf{r+{\bm\delta}}}
M^{BA}_{\mu\nu}({\bm\delta})
t^{\prime\phantom{\dagger}}_{\nu,\mathbf{r}}
\nonumber\\
&+
\sum_{\mathbf{r}\in\Lambda_B}
\sum_{{\bm\delta}}
\sum_{\mu,\nu=x,y,z}
t^{\prime\dagger}_{\mu,\mathbf{r+{\bm\delta}}}
M^{AB}_{\mu\nu}({\bm\delta})
t^{\prime\phantom{\dagger}}_{\nu,\mathbf{r}}
\end{align}
describes the hopping of triplets, where $\Lambda_A$ ($\Lambda_B$) denotes the lattice of A  (B) dimers, and the sum $\sum_{{\bm\delta}}$ is over the four ${\bm\delta}$ nearest neighbor unit vectors $\pm\mathbf{\hat x}=\pm(1,0)$ and $\pm\mathbf{\hat y}=\pm(0,1)$. The matrices $M$ are given as 
\begin{align}
\label{eq:MBA}
M^{BA}(\pm \mathbf{\hat x}) =& \frac{1}{2}
\left(
\begin{array}{ccc}
 0 & -D'_\bot  & 0  \\
 D'_\bot & 0 & \pm \tilde D'_\parallel  \\
0 & \mp \tilde D'_\parallel  & 0
\end{array}\right),
\nonumber\\
M^{BA}(\pm \mathbf{\hat y}) =& \frac{1}{2}
\left(
\begin{array}{ccc}
 0 & -D'_\bot  & \mp \tilde D'_\parallel \\
D'_\bot & 0 & 0  \\
 \pm \tilde D'_\parallel & 0 & 0
\end{array}\right)  
\end{align}
for triplons hopping from A to B dimers and
\begin{align}
M^{AB}(\pm \mathbf{\hat x}) =& \frac{1}{2}
\left(
\begin{array}{ccc}
 0 & D'_\bot & 0  \\
 -D'_\bot & 0 & \pm \tilde D'_\parallel  \\
 0 & \mp \tilde D'_\parallel & 0
\end{array}\right) ,
\nonumber\\
M^{AB}(\pm \mathbf{\hat y}) =& \frac{1}{2}
\left(
\begin{array}{ccc}
 0 & D'_\bot  & \mp \tilde D'_\parallel \\
 -D'_\bot & 0 & 0  \\
\pm \tilde D'_\parallel & 0 & 0
\end{array}\right) 
\end{align}
for  triplons hopping from B to A dimers. Here
\begin{equation}
\tilde D'_\parallel = \left(D'_{\parallel,s} - \frac{ D J'}{2J}\right) .
\end{equation}
Note that $D'_{\parallel,ns}$, the non-staggered component of the inter-dimer DM vector, does not enter above.
We can introduce a unitary transformation for the triplons on the B sublattice that renders the Hamiltonian (\ref{eq:Hamilton}) fully translationally invariant with a single dimer in the unit cell:
\begin{equation}
\mathbf{\tilde t^{\phantom{\dagger}}_{\mathbf{r}}} 
= 
\left\{
\begin{array}{ll}
\mathbf{ t^{\prime\phantom{\dagger}}_{\mathbf{r}}}, &\quad \text{if $\mathbf{r}\in\Lambda_A$\,;}\\
&\\
\mathbf{U} \cdot \mathbf{ t^{\prime\phantom{\dagger}}_{\mathbf{r}}}, &\quad \text{if $\mathbf{r}\in\Lambda_B$\,.}
\end{array}
\right.
\end{equation}
The translation invariance condition $\mathbf{M}^{AB}({\bm\delta}) \mathbf{U^\dagger} =
\mathbf{U} \mathbf{M}^{BA}({\bm\delta})  = \mathbf{\tilde M}({\bm\delta})$ is satisfied with
\begin{equation}
 \mathbf{U} = \left(
\begin{array}{ccc}
-i & 0 & 0 \\
0 & -i & 0 \\
0 & 0 & i
\end{array}
\right) \;,
\end{equation}
so that
\begin{align}
\label{eq:Hamilton_realspace}
\mathcal{H}_{\text{hop}}=&
\sum_{\mathbf{r}\in\Lambda}
\sum_{{\bm\delta}}
\sum_{\mu,\nu=x,y,z}
\tilde{t}^{\dagger}_{\mu,\mathbf{r+{\bm\delta}}}
\tilde M_{\mu\nu}({\bm\delta})
\tilde{t}^{\phantom{\dagger}}_{\nu,\mathbf{r}} \;,
\end{align}
with $\mathbf{\tilde M}({\bm\delta})$ given as 
\begin{align}
\tilde M(\pm \mathbf{\hat x}) = & 
\frac{i}{2}
\left(
\begin{array}{ccc}
0 & D'_\bot   & 0 \\
 -D'_\bot  & 0 & \mp\tilde D'_\parallel \\
0 &  \mp\tilde D'_\parallel & 0 
\end{array}
\right), 
\nonumber\\
\tilde M(\pm \mathbf{\hat y}) = &
\frac{i}{2}
\left(
\begin{array}{ccc}
0 & D'_\bot   & \pm\tilde D'_\parallel  \\
-D'_\bot  & 0 & 0 \\
  \pm\tilde D'_\parallel  & 0 & 0 
\end{array}
\right).
\end{align}
Extending the $\mathbf{U}$ to a $4\times4$ matrix to include the $s^\prime_{\mathbf r}$ operators:
\begin{equation}
 \mathbf{U} = \left(
\begin{array}{cccc}
1 & 0 & 0 & 0 \\
0 & -i & 0 & 0 \\
0 & 0 & -i & 0 \\
0 & 0 & 0 & i
\end{array}
\right) \;,
\end{equation}
the $W$ matrices in the main text are then
\begin{align}
W_{\sf h} = \mathbf{V}_A \quad \text{and} \quad W_{\sf v} = \mathbf{U} \cdot \mathbf{V}_B \;.
\end{align}

\subsection{\\ \vspace{1ex}3. Generalized Dirac Cone physics with spin-L matrices}

Let us consider a generalization of the usual two-band Hamiltonian to one involving spin-$L$ matrices:
\begin{equation}
\mathcal{H} = \sum_\mathbf{k} \mathbf{d} (\mathbf{k}) \cdot \mathbf{L} \;,
\label{Eq.dQ}
\end{equation}
where $\mathbf{d}(k)$ is a fictitious magnetic field and $\mathbf{L}=(L^x,L^y,L^z)$ are $(2L+1)\times(2L+1)$ matrices  satisfying $[L^x,L^y]=iL^z$, $[L^z,L^x]=iL^y$, and $[L^y,L^z]=iL^x$.  For spin-1/2, we recover the usual two-band physics with Dirac cones; for spin-1, we recover the triplon Hamiltonian for SrCu$_2$(BO$_3$)$_2$ defined in the main text. In general, at each $\mathbf{k}$, we have eigenvalues 
\begin{equation}
 \omega_{n}= n d(\mathbf{k}) \;,
\end{equation} 
where $n = -L,\dots,L$ and $d(\mathbf{k})=|\mathbf{d} (\mathbf{k})|$. As $\mathbf{k}$ is continuously varied over the BZ, each of these eigenvalues can be taken to form a band. We have $(2L+1)$ such bands, with the spacing between each pair of bands given by $d(\mathbf{k})$.
If the amplitude of $\mathbf{d}$ is never zero over the BZ, no two bands touch, they are well-separated that can be indexed by $n$. 

Before we evaluate the Chern numbers of these bands, we define the skyrmion number of the $\mathbf{d}$ field:
\begin{equation}
N_s = \frac{1}{4\pi}\int_{\text{BZ}} \! dk_x dk_y \;
 \mathbf{\hat{d}}(\mathbf{k}) \cdot 
 \left[ \frac{\partial \mathbf{\hat{d}}(\mathbf{k})}{\partial k_y} \times \frac{\partial \mathbf{\hat{d}}(\mathbf{k})}{\partial k_x}
\right]\;,
\label{Eq.Ns}
\end{equation}
where $\mathbf{\hat{d}}(\mathbf{k})=\mathbf{d}(\mathbf{k})/d(\mathbf{k})$ is a unit vector.

The Berry curvature of a band $n$ is given by\cite{Bernevig}
\begin{align}
F^{xy}_{n} (\mathbf{k}) = 2i \sum_{m\neq n } \mathrm{Im} 
\frac{ 
\langle n \vert \frac{\partial \mathcal{H}}{\partial k_x} \vert m\rangle \langle m \vert \frac{\partial \mathcal{H}}{\partial k_y} \vert n \rangle }{(\omega_n-\omega_m)^2} \;,
 \label{eq:Fxydef}
\end{align}
where 
\begin{align}
  \frac{\partial \mathcal{H}}{\partial k_x} &= \frac{\partial \mathbf{d}}{\partial k_x}\cdot \mathbf{L} \equiv \mathbf{f}(\mathbf{k}) \cdot\mathbf{L} \;,\nonumber \\
  \frac{\partial \mathcal{H}}{\partial k_y} &= \frac{\partial \mathbf{d}}{\partial k_y}\cdot \mathbf{L} \equiv \mathbf{g}(\mathbf{k}) \cdot\mathbf{L}  \;.
  \label{eq:fgdef}
\end{align}
We use $\mathbf{f}(\mathbf{k})$ and $\mathbf{g}(\mathbf{k})$ as a shorthand notation for $\partial \mathbf{d}/\partial k_x$ and $\partial \mathbf{d}/\partial k_y$, respectively. 

To evaluate the matrix elements in Eq.~(\ref{eq:Fxydef}), we define  a local coordinate system so that the local $z$ axis points along $\mathbf{d}(\mathbf{k})$. We use the form of $\mathcal{H}$, Eq.~(\ref{Eq.dQ}), and Eq.~(\ref{eq:fgdef}) to write
\begin{widetext}
\begin{align}
F^{xy}_{n} (\mathbf{k}) = 2i \sum_{m\neq n } \mathrm{Im} \frac{ \langle n \vert (f^z L^z  + f^- L^+ + f^+ L^-) \vert m\rangle \langle m \vert (g^z L^z  + g^- L^+ + g^+ L^-) \vert n \rangle }{(\omega_n-\omega_m)^2}.
\end{align}
The operators $L^{z}$, $L^{+}$, and $L^{-}$ are spin operators in the local basis, with $z$ along $\mathbf{d}(\mathbf{k})$. They have the usual matrix elements for spin operators. Furthermore, $f^\pm = (f_x \pm i f_y)/2$ and $g^\pm = (g_x \pm i g_y)/2$.
In the expression for the Berry curvature, the only intermediate states that contribute are the immediately higher and lower states $m=n\pm 1$: 
\begin{align}
F^{xy}_{n} (\mathbf{k}) 
&= \frac{2i }{d^2} \mathrm{Im} \left( f^- g^+
 \langle n \vert L^+ \vert n\!-\!1\rangle \langle n\!-\!1 \vert L^- \vert n \rangle 
+ f^+ g^- \langle n \vert L^- \vert n\!+\!1\rangle \langle n\!+\!1 \vert L^+ \vert n \rangle
\right) 
\nonumber\\
&= \frac{2i }{d^2} \mathrm{Im} \left( f^- g^+
 \langle n \vert L^+ L^- \vert n \rangle 
+ f^+ g^- \langle n \vert L^- L^+ \vert n \rangle
\right) ,
 \label{eq:SFtmp5}
\end{align}
\end{widetext}
where we have used the fact that the energy difference $\omega_n-\omega_{n\pm 1}$ is simply $\mp d$, the amplitude of the $\mathbf{d}(\mathbf{k})$ vector.
Next, we use the commutation relation $L^+ L^- - L^- L^+ =  2 L^z$ and
\begin{align}
 \mathrm{Im} f^- g^+  
= - \mathrm{Im} f^+ g^-
 = \frac{1}{4} (f_x g_y - f_y g_x)  
\end{align}
in Eq.~(\ref{eq:SFtmp5}) to get
\begin{align}
F^{xy}_{n} (\mathbf{k}) &= \frac{i }{d^2}  (f_x g_y - f_y g_x) n  \;.
\end{align}
The quantity $f_x g_y - f_y g_x$ is the $z$-component of $\mathbf{g} \times \mathbf{f}$ in the local coordinate system and can be rewritten as
\begin{align}
F^{xy}_{n} (\mathbf{k}) =  \frac{i}{d^3} n  
(d \mathbf{\hat{z}}) \cdot ( \mathbf{g} \times \mathbf{f} ) \;.
\end{align}
Let us now transform back to the global $(x,y,z)$ coordinate system. The Berry curvature expression has a geometric form; as it denotes the volume of a parallelepiped, it does not change under rotation of the coordinate system. Therefore, we have
\begin{align}
F^{xy}_{n} (\mathbf{k}) 
&=  \frac{i}{d^3} n 
\mathbf{d} \cdot \left( \frac{\partial \mathbf{d}}{\partial k_y} \times \frac{\partial \mathbf{d}}{\partial k_x} \right) 
\label{eq:SFddd}\\
&=  i n 
\mathbf{\hat{d}} \cdot \left( \frac{\partial \mathbf{\hat{d}}}{\partial k_y} \times \frac{\partial \mathbf{\hat{d}}}{\partial k_x} \right) \;, 
\label{eq:SFhdhdhd}
\end{align}
where we have used
\begin{equation}
  \frac{\partial \mathbf{\hat{d}}}{\partial k_\mu} =  
  \frac{1}{d} \frac{\partial \mathbf{d}}{\partial k_\mu} 
  - \frac{\mathbf{d}}{d^2} \frac{\partial d}{\partial k_\mu} \;,
\end{equation}
with $\mu=x,y$. Now, the Chern number of the band $n$ is given by
\begin{align}
C_{n} = \frac{1}{2\pi i} \int_{\text{BZ}} \! d{k_x} d{k_y} \; F^{xy}_n = 2 n N_{s} \;,
\end{align}
where $N_s$, defined in  Eq.~(\ref{Eq.Ns}), measures the number of skyrmions in the $\mathbf{d}$ field. We show some examples in Tab.~\ref{tab:cherns}. For the case of $L=1$ matrices as in SrCu$_2$(BO$_3$)$_2$, we have three bands with Chern numbers $2 N_s$, 0 and $-2N_s$.

\begin{table}[t]
\begin{center}
\caption{\label{tab:cherns}The Chern number of the $n^\text{th}$ band for $L=1/2$, $1$, and $3/2$.}
\begin{tabular}{c|ccc}
\hline
\hline
~$n$~ & ~$L=1/2$~ & ~$L =1$~ & ~$L = 3/2$~\\
\hline
$3/2$ & & & $3 N_s$\\
$1$ & &$2 N_s$&\\
$1/2$ & $N_s$& & $N_s$\\
$0$ & & $0$& \\
$-1/2$ & $-N_s$ & & $-N_s$\\
$-1$ & &$-2 N_s$& \\
$-3/2$& & & $-3 N_s$ \\
\hline
\hline
\end{tabular} 
\end{center}
\end{table}

\subsection{\\ \vspace{1ex}4. Thermal Hall conductivity}

We apply the expression for thermal conductivity 
\begin{align}
 \kappa^{xy} &=
  \frac{1}{\beta} 
  \sum_{n=-L}^{L}
  \int_{\text{BZ}} d^2\mathbf k \; c_2(\rho_n) 2 \,\text{Im} \left\langle \partial_{k_x} \psi_n | \partial_{k_y} \psi_n \right\rangle \nonumber\\
  &=  \frac{1}{\beta} \sum_{n=-L}^{L}\int_{\text{BZ}} d^2\mathbf k \; c_2(\rho_n) (-i) F^{xy}_{n}(\mathbf k) \;,
  \label{eq:kappa}
\end{align}
derived by Matsumoto {\it et al.}, to the case of SrCu$_2$(BO$_3$)$_2$. $\beta=1/k_\text{B} T$ is the inverse temperature and
\begin{align}
  \rho_n &= \frac{1}{e^{\omega_n\beta}-1}\;, \nonumber\\
  c_2(\rho) &= \int_{0}^{\rho}d t \; \ln^2(1+t^{-1})\;.
\end{align}
Triplons in SrCu$_2$(BO$_3$)$_2$ are described by the Hamiltonian in Eq.~(\ref{Eq.dQ}) with $L=1$ matrices. There are three bands corresponding to $n=-1,0,1$. Following Eq.~(\ref{eq:SFddd}), the central band with $n=0$ has zero Berry curvature. The upper and lower bands have Berry curvatures with opposite signs, $ F_{+1}^{xy}(\mathbf{k}) = -F_{-1}^{xy}(\mathbf{k}) $. The expression for the thermal Hall effect simplifies to 
\begin{align}
 \kappa^{xy} &=  \int_{\text{BZ}} d^2\mathbf k \; \frac{c_2(\rho_{+1})-c_2(\rho_{-1})}{\beta}  
\frac{\mathbf{d}}{d^3} 
 \cdot \left( \frac{\partial \mathbf{d}}{\partial k_y} \times \frac{\partial \mathbf{d}}{\partial k_x} \right)  \;.
 \label{eq:kappaxydiff}
\end{align}
Since the band dispersions and splittings are much smaller than the gap between the bands and the ground state, $\omega_{+1}-\omega_{-1}\ll \omega_{n}$, we expand Eq.~(\ref{eq:kappaxydiff}) in $d/\omega_{0}$: 
\begin{align}
 c_2(\rho_{+1})-c_2(\rho_{-1}) &= \int_{\rho_{-1}}^{\rho_{+1}}d t \; \ln^2(1+t^{-1})
 \nonumber\\
 &\approx \left(\rho_{+1}-\rho_{-1}\right) \ln^2(1+\rho_{0}^{-1}) 
 \nonumber\\
 &\approx \left(\rho_{+1}-\rho_{-1}\right)  (\omega_0\beta)^2 \;,
\end{align}
where the difference of Bose occupation numbers is
\begin{align}
\rho_{+1}-\rho_{-1} &= 
\frac{1}{e^{\omega_{+1}\beta}-1}-\frac{1}{e^{\omega_{-1}\beta}-1} \nonumber\\
&=-\frac{d \beta}{2\sinh^2(\frac{\omega_0\beta}{2})}+O\left(d^3\right)\;,
\label{eq:rho1m1}
\end{align}
so that 
\begin{align}
\frac{1}{\beta} \left[c_2(\rho_{+1})-c_2(\rho_{-1})\right]  &\approx -\frac{d (\omega_0\beta)^2}{2\sinh^2(\frac{\omega_0\beta}{2})} \;.
  \label{eq:rho1m12}
\end{align}
Eventually, we get the following simple expression for the thermal Hall conductivity:
\begin{equation}
  \kappa^{xy} = R(\omega_0\beta) \kappa^{xy}_\infty \;,
  \label{eq:Romega}
\end{equation}
where  
\begin{equation}
R(x) = \left(\frac{x}{2 \sinh \frac{x}{2}}\right)^2
\end{equation}
and
\begin{equation}
\kappa^{xy}_\infty =  \int_{\text{BZ}} d^2\mathbf k\;  
\frac{2}{d^2}\mathbf{d}\cdot \left( \frac{\partial \mathbf{d}}{\partial k_x} \times \frac{\partial \mathbf{d}}{\partial k_y} \right) \;.
\end{equation}
The temperature dependence stems purely from $R(\omega_0\beta)$. At large temperatures ($\beta \ll \omega_0$), $R(\omega_0\beta)\rightarrow 1$: the conductivity saturates, with $\kappa^{xy}_\infty$ being the high temperature value.
If we increase the temperature from zero, $\kappa^{xy}$ reaches half of its saturation value already at $k_{\text{B}}T \approx \omega_0/3$, where the boson occupation number is $\rho_0 \approx 0.052$. However, recent neutron data on SrCu$_2$(BO$_3$)$_2$\cite{Ronnow2014} suggests that triplon-triplon interactions cannot be neglected for $k_{\text{B}}T \approx \omega_0/3\sim 10K$. We suggest transport measurements focussing on $T\lesssim 5K$ where the neutron intensity shows no damping\cite{Gaulin2004} and a strong thermal Hall effect may be expected.

Note that thermal Hall conductivity can be finite even for magnetic fields $|h|>h_c$ as seen in Fig.~5 of the main text. The Chern numbers are then zero as the integral of the Berry curvature over each band is zero. However, due to the thermal occupation of bosons, different parts of the bands contribute differently to the Hall conductivity, giving a net non-zero $\kappa^{xy}$.

To get a quantitative value for the $\kappa^{xy}$, we numerically integrate Eq.~(\ref{eq:kappa}) using the exchange parameters and $g$-tensor value that reproduce the ESR spectrum at low magnetic fields (see main text).
Using these values,
we obtain $\kappa^{xy}$ in GHz. To convert to the experimentally relevant units of W/(Km), we first multiply by
the Boltzmann factor $k_B$ to obtain JK$^{-1}$s$^{-1}$=WK$^{-1}$. SrCu$_2$(BO$_3$)$_2$ is a layered material with a layer thickness of 0.332 nm. We multiply the single-layer contribution by the number of the layers in a sample of height 1 meter, to obtain $\kappa^{xy}$ with dimensions of WK$^{-1}$m$^{-1}$.

\subsection{\\ \vspace{1ex}5. Role  of second neighbor dimer-dimer exchange}

\begin{figure}[h!]
\begin{center}
\includegraphics[width=0.8\columnwidth]{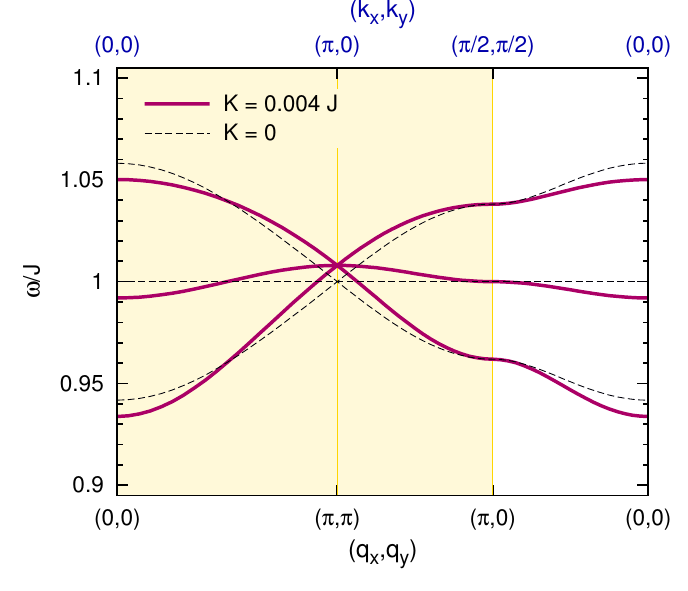}
\caption{
Effect of next nearest triplon hopping $K$ on the spectrum, with $K/J =0.004$, the value extracted from series expansion~\cite{Weihong1999}. 
The black dashed line represents the triplon energies without $K$. We show two horizontal axes; the bottom axis is used in earlier works such as Refs.~\cite{Cheng2007,Gaulin2004} and corresponds to the original (smaller) BZ, the top blue axis shows the $(k_x,k_y)$ components in the extended BZ picture used in this work. The `diagonal' triplet hopping, $K\sim (J'/J)^6$, removes the reflection symmetry of the spectrum about $\omega_0$ and the middle $n=0$ flat mode acquires a dispersion. Including $K$ gives better agreement with neutron spectra~\cite{Gaulin2004} measured along ${\bf q}=(q_x,0)$ (white area).}
\label{fig:j2_effect}
\end{center}
\end{figure}

\begin{figure}[h!]
\begin{center}
\includegraphics[width=0.8\columnwidth]{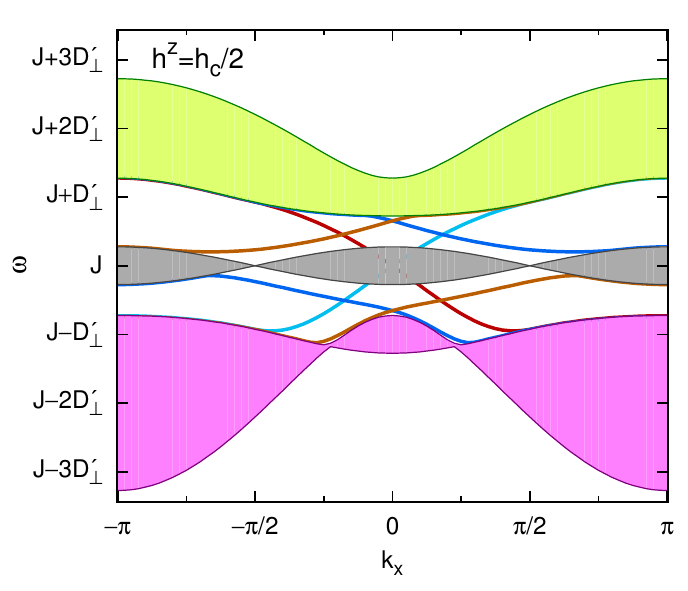}
\caption{Band structure for a strip with $K/J=0.004$. The  middle band acquires a dispersion and the lower and upper Chern bands are deformed, but the topological properties survive along with their fingerprint for open boundaries: the four edge states connecting the Chern bands.
}
\label{fig:open}
\end{center}
\end{figure}

The dispersion of the bands is only partly due to DM interaction. Another, comparable effect comes from the 6$^{\text{th}}$ order perturbation in $J'/J$, as noted in Ref.~\cite{Miyahara1999}. 
The inclusion of this term modifies the hopping matrix $M(\bf{ k})$ of Eq. (5) in main text to 
\begin{equation}
M({\mathbf{k}}) =  \left[J - 2K \gamma_4({\mathbf{k}})\right] \mathbf{1} + \mathbf{d}({\mathbf{k}})\cdot \mathbf{L} \;,
\label{Eq.Hmodform}
\end{equation}
where 
$ \gamma_4({\mathbf{k}}) = \cos k_x \cos k_y $.
The second neighbor dimer-dimer interaction $K$ only appears in the diagonal of $M(\bf{k})$ and does not change the ${\bf d}({\bf k})$ vector. It locally shifts the triplon energies by $-2K\cos k_x\cos k_y$, as shown in Fig.~\ref{fig:j2_effect}. However, it has no effect on topological properties, as the Berry curvature $F^{xy}_n(\bf{k})$ -- which decides Chern numbers and the thermal Hall signal $\kappa^{xy}$ -- only depends on ${\bf d}({\bf k})$. In Fig.~\ref{fig:open},
we show the edge state spectrum on a strip geometry upon including a finite $K$.

The thermal Hall signal $\kappa^{xy}$ does not show any visible changes upon the inclusion of $K$. 
This can be rationalized starting from Eq.~(\ref{eq:kappaxydiff}). This equation still holds as $F^{xy}_0(\bf{k})$ remains $0$ and $F^{xy}_1(\mathbf{k})=-F^{xy}_{-1}(\mathbf{k})$ since ${\bf d}(\mathbf{k})$ and the eigenfunctions do not change. 
The derivation through Eqs.~(\ref{eq:rho1m1}) -- (\ref{eq:rho1m12}) does not change either, since $\omega_{1}$ and $\omega_{-1}$ are shifted by the same value. Therefore we can write
\begin{equation}
\kappa^{xy}=\int_{\text BZ}d^2{\bf k} \frac{(\omega_0(\mathbf{k})\beta)^2}{2 \sinh^2(\frac{\omega_0(\mathbf{k})\beta}{2})}\frac{{\bf d}}{d^2}\cdot\left( \frac{\partial \mathbf{d}}{\partial k_y} \times \frac{\partial \mathbf{d}}{\partial k_x} \right).
\end{equation}
However, since the $\omega_0(\mathbf{k}) = J-2K\gamma_4(\mathbf{k})$ now depends on $\mathbf{k}$ due to the second neighbor triplet hopping, we can no longer factorize $\kappa^{xy}$ as in Eq.~(\ref{eq:Romega}). By performing an expansion in $K/J$, we find a small correction to $\kappa^{xy}$ of order $K/J$.


\begin{thebibliography}{46}%
\makeatletter
\providecommand \@ifxundefined [1]{%
 \@ifx{#1\undefined}
}%
\providecommand \@ifnum [1]{%
 \ifnum #1\expandafter \@firstoftwo
 \else \expandafter \@secondoftwo
 \fi
}%
\providecommand \@ifx [1]{%
 \ifx #1\expandafter \@firstoftwo
 \else \expandafter \@secondoftwo
 \fi
}%
\providecommand \natexlab [1]{#1}%
\providecommand \enquote  [1]{``#1''}%
\providecommand \bibnamefont  [1]{#1}%
\providecommand \bibfnamefont [1]{#1}%
\providecommand \citenamefont [1]{#1}%
\providecommand \href@noop [0]{\@secondoftwo}%
\providecommand \href [0]{\begingroup \@sanitize@url \@href}%
\providecommand \@href[1]{\@@startlink{#1}\@@href}%
\providecommand \@@href[1]{\endgroup#1\@@endlink}%
\providecommand \@sanitize@url [0]{\catcode `\\12\catcode `\$12\catcode
  `\&12\catcode `\#12\catcode `\^12\catcode `\_12\catcode `\%12\relax}%
\providecommand \@@startlink[1]{}%
\providecommand \@@endlink[0]{}%
\providecommand \url  [0]{\begingroup\@sanitize@url \@url }%
\providecommand \@url [1]{\endgroup\@href {#1}{\urlprefix }}%
\providecommand \urlprefix  [0]{URL }%
\providecommand \Eprint [0]{\href }%
\providecommand \doibase [0]{http://dx.doi.org/}%
\providecommand \selectlanguage [0]{\@gobble}%
\providecommand \bibinfo  [0]{\@secondoftwo}%
\providecommand \bibfield  [0]{\@secondoftwo}%
\providecommand \translation [1]{[#1]}%
\providecommand \BibitemOpen [0]{}%
\providecommand \bibitemStop [0]{}%
\providecommand \bibitemNoStop [0]{.\EOS\space}%
\providecommand \EOS [0]{\spacefactor3000\relax}%
\providecommand \BibitemShut  [1]{\csname bibitem#1\endcsname}%
\let\auto@bib@innerbib\@empty
\bibitem [{\citenamefont {Raghu}\ and\ \citenamefont
  {Haldane}(2008)}]{Raghu2008}%
  \BibitemOpen
  \bibfield  {author} {\bibinfo {author} {\bibfnamefont {S.}~\bibnamefont
  {Raghu}}\ and\ \bibinfo {author} {\bibfnamefont {F.~D.~M.}\ \bibnamefont
  {Haldane}},\ }\href {\doibase 10.1103/PhysRevA.78.033834} {\bibfield
  {journal} {\bibinfo  {journal} {Phys. Rev. A}\ }\textbf {\bibinfo {volume}
  {78}},\ \bibinfo {pages} {033834} (\bibinfo {year} {2008})}\BibitemShut
  {NoStop}%
\bibitem [{\citenamefont {Petrescu}\ \emph {et~al.}(2012)\citenamefont
  {Petrescu}, \citenamefont {Houck},\ and\ \citenamefont
  {Le~Hur}}]{Petrescu2012}%
  \BibitemOpen
  \bibfield  {author} {\bibinfo {author} {\bibfnamefont {A.}~\bibnamefont
  {Petrescu}}, \bibinfo {author} {\bibfnamefont {A.~A.}\ \bibnamefont {Houck}},
  \ and\ \bibinfo {author} {\bibfnamefont {K.}~\bibnamefont {Le~Hur}},\ }\href
  {\doibase 10.1103/PhysRevA.86.053804} {\bibfield  {journal} {\bibinfo
  {journal} {Phys. Rev. A}\ }\textbf {\bibinfo {volume} {86}},\ \bibinfo
  {pages} {053804} (\bibinfo {year} {2012})}\BibitemShut {NoStop}%
\bibitem [{\citenamefont {Rechtsman}\ \emph {et~al.}(2013)\citenamefont
  {Rechtsman}, \citenamefont {Zeuner}, \citenamefont {Plotnik}, \citenamefont
  {Lumer}, \citenamefont {Podolsky}, \citenamefont {Dreisow}, \citenamefont
  {Nolte}, \citenamefont {Segev},\ and\ \citenamefont
  {Szameit}}]{Rechtsman2013}%
  \BibitemOpen
  \bibfield  {author} {\bibinfo {author} {\bibfnamefont {M.~C.}\ \bibnamefont
  {Rechtsman}}, \bibinfo {author} {\bibfnamefont {J.~M.}\ \bibnamefont
  {Zeuner}}, \bibinfo {author} {\bibfnamefont {Y.}~\bibnamefont {Plotnik}},
  \bibinfo {author} {\bibfnamefont {Y.}~\bibnamefont {Lumer}}, \bibinfo
  {author} {\bibfnamefont {D.}~\bibnamefont {Podolsky}}, \bibinfo {author}
  {\bibfnamefont {F.}~\bibnamefont {Dreisow}}, \bibinfo {author} {\bibfnamefont
  {S.}~\bibnamefont {Nolte}}, \bibinfo {author} {\bibfnamefont
  {M.}~\bibnamefont {Segev}}, \ and\ \bibinfo {author} {\bibfnamefont
  {A.}~\bibnamefont {Szameit}},\ }\href {\doibase doi:10.1038/nature12066}
  {\bibfield  {journal} {\bibinfo  {journal} {Nature}\ }\textbf {\bibinfo
  {volume} {496}},\ \bibinfo {pages} {196} (\bibinfo {year}
  {2013})}\BibitemShut {NoStop}%
\bibitem [{\citenamefont {Hafezi}\ \emph {et~al.}(2013)\citenamefont {Hafezi},
  \citenamefont {Mittal}, \citenamefont {Fan}, \citenamefont {Migdall},\ and\
  \citenamefont {Taylor}}]{Hafezi2013}%
  \BibitemOpen
  \bibfield  {author} {\bibinfo {author} {\bibfnamefont {M.}~\bibnamefont
  {Hafezi}}, \bibinfo {author} {\bibfnamefont {S.}~\bibnamefont {Mittal}},
  \bibinfo {author} {\bibfnamefont {J.}~\bibnamefont {Fan}}, \bibinfo {author}
  {\bibfnamefont {A.}~\bibnamefont {Migdall}}, \ and\ \bibinfo {author}
  {\bibfnamefont {J.~M.}\ \bibnamefont {Taylor}},\ }\href {\doibase
  doi:10.1038/nphoton.2013.274} {\bibfield  {journal} {\bibinfo  {journal}
  {Nature Photonics}\ }\textbf {\bibinfo {volume} {7}},\ \bibinfo {pages}
  {1001} (\bibinfo {year} {2013})}\BibitemShut {NoStop}%
\bibitem [{\citenamefont {Katsura}\ \emph {et~al.}(2010)\citenamefont
  {Katsura}, \citenamefont {Nagaosa},\ and\ \citenamefont {Lee}}]{Katsura2010}%
  \BibitemOpen
  \bibfield  {author} {\bibinfo {author} {\bibfnamefont {H.}~\bibnamefont
  {Katsura}}, \bibinfo {author} {\bibfnamefont {N.}~\bibnamefont {Nagaosa}}, \
  and\ \bibinfo {author} {\bibfnamefont {P.~A.}\ \bibnamefont {Lee}},\ }\href
  {\doibase 10.1103/PhysRevLett.104.066403} {\bibfield  {journal} {\bibinfo
  {journal} {Phys. Rev. Lett.}\ }\textbf {\bibinfo {volume} {104}},\ \bibinfo
  {pages} {066403} (\bibinfo {year} {2010})}\BibitemShut {NoStop}%
\bibitem [{\citenamefont {Shindou}\ \emph {et~al.}(2013)\citenamefont
  {Shindou}, \citenamefont {Matsumoto}, \citenamefont {Murakami},\ and\
  \citenamefont {Ohe}}]{Shindou2013}%
  \BibitemOpen
  \bibfield  {author} {\bibinfo {author} {\bibfnamefont {R.}~\bibnamefont
  {Shindou}}, \bibinfo {author} {\bibfnamefont {R.}~\bibnamefont {Matsumoto}},
  \bibinfo {author} {\bibfnamefont {S.}~\bibnamefont {Murakami}}, \ and\
  \bibinfo {author} {\bibfnamefont {J.-i.}\ \bibnamefont {Ohe}},\ }\href
  {\doibase 10.1103/PhysRevB.87.174427} {\bibfield  {journal} {\bibinfo
  {journal} {Phys. Rev. B}\ }\textbf {\bibinfo {volume} {87}},\ \bibinfo
  {pages} {174427} (\bibinfo {year} {2013})}\BibitemShut {NoStop}%
\bibitem [{\citenamefont {Matsumoto}\ and\ \citenamefont
  {Murakami}(2011)}]{Matsumoto2011}%
  \BibitemOpen
  \bibfield  {author} {\bibinfo {author} {\bibfnamefont {R.}~\bibnamefont
  {Matsumoto}}\ and\ \bibinfo {author} {\bibfnamefont {S.}~\bibnamefont
  {Murakami}},\ }\href {\doibase 10.1103/PhysRevLett.106.197202} {\bibfield
  {journal} {\bibinfo  {journal} {Phys. Rev. Lett.}\ }\textbf {\bibinfo
  {volume} {106}},\ \bibinfo {pages} {197202} (\bibinfo {year}
  {2011})}\BibitemShut {NoStop}%
\bibitem [{\citenamefont {Ideue}\ \emph {et~al.}(2012)\citenamefont {Ideue},
  \citenamefont {Onose}, \citenamefont {Katsura}, \citenamefont {Shiomi},
  \citenamefont {Ishiwata}, \citenamefont {Nagaosa},\ and\ \citenamefont
  {Tokura}}]{Ideue2012}%
  \BibitemOpen
  \bibfield  {author} {\bibinfo {author} {\bibfnamefont {T.}~\bibnamefont
  {Ideue}}, \bibinfo {author} {\bibfnamefont {Y.}~\bibnamefont {Onose}},
  \bibinfo {author} {\bibfnamefont {H.}~\bibnamefont {Katsura}}, \bibinfo
  {author} {\bibfnamefont {Y.}~\bibnamefont {Shiomi}}, \bibinfo {author}
  {\bibfnamefont {S.}~\bibnamefont {Ishiwata}}, \bibinfo {author}
  {\bibfnamefont {N.}~\bibnamefont {Nagaosa}}, \ and\ \bibinfo {author}
  {\bibfnamefont {Y.}~\bibnamefont {Tokura}},\ }\href {\doibase
  10.1103/PhysRevB.85.134411} {\bibfield  {journal} {\bibinfo  {journal} {Phys.
  Rev. B}\ }\textbf {\bibinfo {volume} {85}},\ \bibinfo {pages} {134411}
  (\bibinfo {year} {2012})}\BibitemShut {NoStop}%
\bibitem [{\citenamefont {Zhang}\ \emph {et~al.}(2013)\citenamefont {Zhang},
  \citenamefont {Ren}, \citenamefont {Wang},\ and\ \citenamefont
  {Li}}]{Zhang2013}%
  \BibitemOpen
  \bibfield  {author} {\bibinfo {author} {\bibfnamefont {L.}~\bibnamefont
  {Zhang}}, \bibinfo {author} {\bibfnamefont {J.}~\bibnamefont {Ren}}, \bibinfo
  {author} {\bibfnamefont {J.-S.}\ \bibnamefont {Wang}}, \ and\ \bibinfo
  {author} {\bibfnamefont {B.}~\bibnamefont {Li}},\ }\href {\doibase
  10.1103/PhysRevB.87.144101} {\bibfield  {journal} {\bibinfo  {journal} {Phys.
  Rev. B}\ }\textbf {\bibinfo {volume} {87}},\ \bibinfo {pages} {144101}
  (\bibinfo {year} {2013})}\BibitemShut {NoStop}%
\bibitem [{\citenamefont {Zhang}\ \emph {et~al.}(2010)\citenamefont {Zhang},
  \citenamefont {Ren}, \citenamefont {Wang},\ and\ \citenamefont
  {Li}}]{Zhang2010}%
  \BibitemOpen
  \bibfield  {author} {\bibinfo {author} {\bibfnamefont {L.}~\bibnamefont
  {Zhang}}, \bibinfo {author} {\bibfnamefont {J.}~\bibnamefont {Ren}}, \bibinfo
  {author} {\bibfnamefont {J.-S.}\ \bibnamefont {Wang}}, \ and\ \bibinfo
  {author} {\bibfnamefont {B.}~\bibnamefont {Li}},\ }\href {\doibase
  10.1103/PhysRevLett.105.225901} {\bibfield  {journal} {\bibinfo  {journal}
  {Phys. Rev. Lett.}\ }\textbf {\bibinfo {volume} {105}},\ \bibinfo {pages}
  {225901} (\bibinfo {year} {2010})}\BibitemShut {NoStop}%
\bibitem [{\citenamefont {Zhang}\ \emph {et~al.}(2011)\citenamefont {Zhang},
  \citenamefont {Ren}, \citenamefont {Wang},\ and\ \citenamefont
  {Li}}]{Zhang2011}%
  \BibitemOpen
  \bibfield  {author} {\bibinfo {author} {\bibfnamefont {L.}~\bibnamefont
  {Zhang}}, \bibinfo {author} {\bibfnamefont {J.}~\bibnamefont {Ren}}, \bibinfo
  {author} {\bibfnamefont {J.-S.}\ \bibnamefont {Wang}}, \ and\ \bibinfo
  {author} {\bibfnamefont {B.}~\bibnamefont {Li}},\ }\href
  {http://stacks.iop.org/0953-8984/23/i=30/a=305402} {\bibfield  {journal}
  {\bibinfo  {journal} {Journal of Physics: Condensed Matter}\ }\textbf
  {\bibinfo {volume} {23}},\ \bibinfo {pages} {305402} (\bibinfo {year}
  {2011})}\BibitemShut {NoStop}%
\bibitem [{\citenamefont {Qin}\ \emph {et~al.}(2012)\citenamefont {Qin},
  \citenamefont {Zhou},\ and\ \citenamefont {Shi}}]{Qin2012}%
  \BibitemOpen
  \bibfield  {author} {\bibinfo {author} {\bibfnamefont {T.}~\bibnamefont
  {Qin}}, \bibinfo {author} {\bibfnamefont {J.}~\bibnamefont {Zhou}}, \ and\
  \bibinfo {author} {\bibfnamefont {J.}~\bibnamefont {Shi}},\ }\href {\doibase
  10.1103/PhysRevB.86.104305} {\bibfield  {journal} {\bibinfo  {journal} {Phys.
  Rev. B}\ }\textbf {\bibinfo {volume} {86}},\ \bibinfo {pages} {104305}
  (\bibinfo {year} {2012})}\BibitemShut {NoStop}%
\bibitem [{\citenamefont {van Hoogdalem}\ \emph {et~al.}(2013)\citenamefont
  {van Hoogdalem}, \citenamefont {Tserkovnyak},\ and\ \citenamefont
  {Loss}}]{Hoogdalem2013}%
  \BibitemOpen
  \bibfield  {author} {\bibinfo {author} {\bibfnamefont {K.~A.}\ \bibnamefont
  {van Hoogdalem}}, \bibinfo {author} {\bibfnamefont {Y.}~\bibnamefont
  {Tserkovnyak}}, \ and\ \bibinfo {author} {\bibfnamefont {D.}~\bibnamefont
  {Loss}},\ }\href {\doibase 10.1103/PhysRevB.87.024402} {\bibfield  {journal}
  {\bibinfo  {journal} {Phys. Rev. B}\ }\textbf {\bibinfo {volume} {87}},\
  \bibinfo {pages} {024402} (\bibinfo {year} {2013})}\BibitemShut {NoStop}%
\bibitem [{\citenamefont {Onose}\ \emph {et~al.}(2010)\citenamefont {Onose},
  \citenamefont {Ideue}, \citenamefont {Katsura}, \citenamefont {Shiomi},
  \citenamefont {Nagaosa},\ and\ \citenamefont {Tokura}}]{Onose2010}%
  \BibitemOpen
  \bibfield  {author} {\bibinfo {author} {\bibfnamefont {Y.}~\bibnamefont
  {Onose}}, \bibinfo {author} {\bibfnamefont {T.}~\bibnamefont {Ideue}},
  \bibinfo {author} {\bibfnamefont {H.}~\bibnamefont {Katsura}}, \bibinfo
  {author} {\bibfnamefont {Y.}~\bibnamefont {Shiomi}}, \bibinfo {author}
  {\bibfnamefont {N.}~\bibnamefont {Nagaosa}}, \ and\ \bibinfo {author}
  {\bibfnamefont {Y.}~\bibnamefont {Tokura}},\ }\href {\doibase
  10.1126/science.1188260} {\bibfield  {journal} {\bibinfo  {journal}
  {Science}\ }\textbf {\bibinfo {volume} {329}},\ \bibinfo {pages} {297}
  (\bibinfo {year} {2010})}\BibitemShut {NoStop}%
\bibitem [{\citenamefont {Shastry}\ and\ \citenamefont
  {Sutherland}(1981)}]{Shastry1981}%
  \BibitemOpen
  \bibfield  {author} {\bibinfo {author} {\bibfnamefont {B.~S.}\ \bibnamefont
  {Shastry}}\ and\ \bibinfo {author} {\bibfnamefont {B.}~\bibnamefont
  {Sutherland}},\ }\href {\doibase
  http://dx.doi.org/10.1016/0378-4363(81)90838-X} {\bibfield  {journal}
  {\bibinfo  {journal} {Physica B+C}\ }\textbf {\bibinfo {volume} {108}},\
  \bibinfo {pages} {1069 } (\bibinfo {year} {1981})}\BibitemShut {NoStop}%
\bibitem [{\citenamefont {Miyahara}\ and\ \citenamefont
  {Ueda}(1999)}]{Miyahara1999}%
  \BibitemOpen
  \bibfield  {author} {\bibinfo {author} {\bibfnamefont {S.}~\bibnamefont
  {Miyahara}}\ and\ \bibinfo {author} {\bibfnamefont {K.}~\bibnamefont
  {Ueda}},\ }\href {\doibase 10.1103/PhysRevLett.82.3701} {\bibfield  {journal}
  {\bibinfo  {journal} {Phys. Rev. Lett.}\ }\textbf {\bibinfo {volume} {82}},\
  \bibinfo {pages} {3701} (\bibinfo {year} {1999})}\BibitemShut {NoStop}%
\bibitem [{\citenamefont {Smith}\ and\ \citenamefont
  {Keszler}(1989)}]{Smith1989}%
  \BibitemOpen
  \bibfield  {author} {\bibinfo {author} {\bibfnamefont {R.~W.}\ \bibnamefont
  {Smith}}\ and\ \bibinfo {author} {\bibfnamefont {D.~A.}\ \bibnamefont
  {Keszler}},\ }\href {\doibase http://dx.doi.org/10.1016/0022-4596(89)90019-4}
  {\bibfield  {journal} {\bibinfo  {journal} {Journal of Solid State
  Chemistry}\ }\textbf {\bibinfo {volume} {81}},\ \bibinfo {pages} {305 }
  (\bibinfo {year} {1989})}\BibitemShut {NoStop}%
\bibitem [{\citenamefont {Kageyama}\ \emph {et~al.}(1999)\citenamefont
  {Kageyama}, \citenamefont {Yoshimura}, \citenamefont {Stern}, \citenamefont
  {Mushnikov}, \citenamefont {Onizuka}, \citenamefont {Kato}, \citenamefont
  {Kosuge}, \citenamefont {Slichter}, \citenamefont {Goto},\ and\ \citenamefont
  {Ueda}}]{Kageyama1999}%
  \BibitemOpen
  \bibfield  {author} {\bibinfo {author} {\bibfnamefont {H.}~\bibnamefont
  {Kageyama}}, \bibinfo {author} {\bibfnamefont {K.}~\bibnamefont {Yoshimura}},
  \bibinfo {author} {\bibfnamefont {R.}~\bibnamefont {Stern}}, \bibinfo
  {author} {\bibfnamefont {N.~V.}\ \bibnamefont {Mushnikov}}, \bibinfo {author}
  {\bibfnamefont {K.}~\bibnamefont {Onizuka}}, \bibinfo {author} {\bibfnamefont
  {M.}~\bibnamefont {Kato}}, \bibinfo {author} {\bibfnamefont {K.}~\bibnamefont
  {Kosuge}}, \bibinfo {author} {\bibfnamefont {C.~P.}\ \bibnamefont
  {Slichter}}, \bibinfo {author} {\bibfnamefont {T.}~\bibnamefont {Goto}}, \
  and\ \bibinfo {author} {\bibfnamefont {Y.}~\bibnamefont {Ueda}},\ }\href
  {\doibase 10.1103/PhysRevLett.82.3168} {\bibfield  {journal} {\bibinfo
  {journal} {Phys. Rev. Lett.}\ }\textbf {\bibinfo {volume} {82}},\ \bibinfo
  {pages} {3168} (\bibinfo {year} {1999})}\BibitemShut {NoStop}%
\bibitem [{\citenamefont {Sachdev}\ and\ \citenamefont
  {Bhatt}(1990)}]{SachdevBhatt1990}%
  \BibitemOpen
  \bibfield  {author} {\bibinfo {author} {\bibfnamefont {S.}~\bibnamefont
  {Sachdev}}\ and\ \bibinfo {author} {\bibfnamefont {R.~N.}\ \bibnamefont
  {Bhatt}},\ }\href {\doibase 10.1103/PhysRevB.41.9323} {\bibfield  {journal}
  {\bibinfo  {journal} {Phys. Rev. B}\ }\textbf {\bibinfo {volume} {41}},\
  \bibinfo {pages} {9323} (\bibinfo {year} {1990})}\BibitemShut {NoStop}%
\bibitem [{\citenamefont {Giamarchi}\ \emph {et~al.}(2008)\citenamefont
  {Giamarchi}, \citenamefont {Ruegg},\ and\ \citenamefont
  {Tchernyshyov}}]{Giamarchi2008}%
  \BibitemOpen
  \bibfield  {author} {\bibinfo {author} {\bibfnamefont {T.}~\bibnamefont
  {Giamarchi}}, \bibinfo {author} {\bibfnamefont {C.}~\bibnamefont {Ruegg}}, \
  and\ \bibinfo {author} {\bibfnamefont {O.}~\bibnamefont {Tchernyshyov}},\
  }\href {\doibase 10.1038/nphys893} {\bibfield  {journal} {\bibinfo  {journal}
  {Nat Phys}\ }\textbf {\bibinfo {volume} {4}},\ \bibinfo {pages} {198}
  (\bibinfo {year} {2008})}\BibitemShut {NoStop}%
\bibitem [{\citenamefont {Momoi}\ and\ \citenamefont
  {Totsuka}(2000)}]{Momoi2000}%
  \BibitemOpen
  \bibfield  {author} {\bibinfo {author} {\bibfnamefont {T.}~\bibnamefont
  {Momoi}}\ and\ \bibinfo {author} {\bibfnamefont {K.}~\bibnamefont
  {Totsuka}},\ }\href {\doibase 10.1103/PhysRevB.62.15067} {\bibfield
  {journal} {\bibinfo  {journal} {Phys. Rev. B}\ }\textbf {\bibinfo {volume}
  {62}},\ \bibinfo {pages} {15067} (\bibinfo {year} {2000})}\BibitemShut
  {NoStop}%
\bibitem [{\citenamefont {Nojiri}\ \emph {et~al.}(2003)\citenamefont {Nojiri},
  \citenamefont {Kageyama}, \citenamefont {Ueda},\ and\ \citenamefont
  {Motokawa}}]{Nojiri2003}%
  \BibitemOpen
  \bibfield  {author} {\bibinfo {author} {\bibfnamefont {H.}~\bibnamefont
  {Nojiri}}, \bibinfo {author} {\bibfnamefont {H.}~\bibnamefont {Kageyama}},
  \bibinfo {author} {\bibfnamefont {Y.}~\bibnamefont {Ueda}}, \ and\ \bibinfo
  {author} {\bibfnamefont {M.}~\bibnamefont {Motokawa}},\ }\href {\doibase
  10.1143/JPSJ.72.3243} {\bibfield  {journal} {\bibinfo  {journal} {Journal of
  the Physical Society of Japan}\ }\textbf {\bibinfo {volume} {72}},\ \bibinfo
  {pages} {3243} (\bibinfo {year} {2003})}\BibitemShut {NoStop}%
\bibitem [{\citenamefont {R{\~o}{\~o}m}\ \emph {et~al.}(2004)\citenamefont
  {R{\~o}{\~o}m}, \citenamefont {H\"uvonen}, \citenamefont {Nagel},
  \citenamefont {Hwang}, \citenamefont {Timusk},\ and\ \citenamefont
  {Kageyama}}]{Room2004}%
  \BibitemOpen
  \bibfield  {author} {\bibinfo {author} {\bibfnamefont {T.}~\bibnamefont
  {R{\~o}{\~o}m}}, \bibinfo {author} {\bibfnamefont {D.}~\bibnamefont
  {H\"uvonen}}, \bibinfo {author} {\bibfnamefont {U.}~\bibnamefont {Nagel}},
  \bibinfo {author} {\bibfnamefont {J.}~\bibnamefont {Hwang}}, \bibinfo
  {author} {\bibfnamefont {T.}~\bibnamefont {Timusk}}, \ and\ \bibinfo {author}
  {\bibfnamefont {H.}~\bibnamefont {Kageyama}},\ }\href {\doibase
  10.1103/PhysRevB.70.144417} {\bibfield  {journal} {\bibinfo  {journal} {Phys.
  Rev. B}\ }\textbf {\bibinfo {volume} {70}},\ \bibinfo {pages} {144417}
  (\bibinfo {year} {2004})}\BibitemShut {NoStop}%
\bibitem [{\citenamefont {Gaulin}\ \emph {et~al.}(2004)\citenamefont {Gaulin},
  \citenamefont {Lee}, \citenamefont {Haravifard}, \citenamefont {Castellan},
  \citenamefont {Berlinsky}, \citenamefont {Dabkowska}, \citenamefont {Qiu},\
  and\ \citenamefont {Copley}}]{Gaulin2004}%
  \BibitemOpen
  \bibfield  {author} {\bibinfo {author} {\bibfnamefont {B.~D.}\ \bibnamefont
  {Gaulin}}, \bibinfo {author} {\bibfnamefont {S.~H.}\ \bibnamefont {Lee}},
  \bibinfo {author} {\bibfnamefont {S.}~\bibnamefont {Haravifard}}, \bibinfo
  {author} {\bibfnamefont {J.~P.}\ \bibnamefont {Castellan}}, \bibinfo {author}
  {\bibfnamefont {A.~J.}\ \bibnamefont {Berlinsky}}, \bibinfo {author}
  {\bibfnamefont {H.~A.}\ \bibnamefont {Dabkowska}}, \bibinfo {author}
  {\bibfnamefont {Y.}~\bibnamefont {Qiu}}, \ and\ \bibinfo {author}
  {\bibfnamefont {J.~R.~D.}\ \bibnamefont {Copley}},\ }\href {\doibase
  10.1103/PhysRevLett.93.267202} {\bibfield  {journal} {\bibinfo  {journal}
  {Phys. Rev. Lett.}\ }\textbf {\bibinfo {volume} {93}},\ \bibinfo {pages}
  {267202} (\bibinfo {year} {2004})}\BibitemShut {NoStop}%
\bibitem [{\citenamefont {Gozar}\ \emph {et~al.}(2005)\citenamefont {Gozar},
  \citenamefont {Dennis}, \citenamefont {Kageyama},\ and\ \citenamefont
  {Blumberg}}]{Gozar2005}%
  \BibitemOpen
  \bibfield  {author} {\bibinfo {author} {\bibfnamefont {A.}~\bibnamefont
  {Gozar}}, \bibinfo {author} {\bibfnamefont {B.~S.}\ \bibnamefont {Dennis}},
  \bibinfo {author} {\bibfnamefont {H.}~\bibnamefont {Kageyama}}, \ and\
  \bibinfo {author} {\bibfnamefont {G.}~\bibnamefont {Blumberg}},\ }\href
  {\doibase 10.1103/PhysRevB.72.064405} {\bibfield  {journal} {\bibinfo
  {journal} {Phys. Rev. B}\ }\textbf {\bibinfo {volume} {72}},\ \bibinfo
  {pages} {064405} (\bibinfo {year} {2005})}\BibitemShut {NoStop}%
\bibitem [{\citenamefont {C\'epas}\ \emph {et~al.}(2001)\citenamefont
  {C\'epas}, \citenamefont {Kakurai}, \citenamefont {Regnault}, \citenamefont
  {Ziman}, \citenamefont {Boucher}, \citenamefont {Aso}, \citenamefont {Nishi},
  \citenamefont {Kageyama},\ and\ \citenamefont {Ueda}}]{Cepas2001}%
  \BibitemOpen
  \bibfield  {author} {\bibinfo {author} {\bibfnamefont {O.}~\bibnamefont
  {C\'epas}}, \bibinfo {author} {\bibfnamefont {K.}~\bibnamefont {Kakurai}},
  \bibinfo {author} {\bibfnamefont {L.~P.}\ \bibnamefont {Regnault}}, \bibinfo
  {author} {\bibfnamefont {T.}~\bibnamefont {Ziman}}, \bibinfo {author}
  {\bibfnamefont {J.~P.}\ \bibnamefont {Boucher}}, \bibinfo {author}
  {\bibfnamefont {N.}~\bibnamefont {Aso}}, \bibinfo {author} {\bibfnamefont
  {M.}~\bibnamefont {Nishi}}, \bibinfo {author} {\bibfnamefont
  {H.}~\bibnamefont {Kageyama}}, \ and\ \bibinfo {author} {\bibfnamefont
  {Y.}~\bibnamefont {Ueda}},\ }\href {\doibase 10.1103/PhysRevLett.87.167205}
  {\bibfield  {journal} {\bibinfo  {journal} {Phys. Rev. Lett.}\ }\textbf
  {\bibinfo {volume} {87}},\ \bibinfo {pages} {167205} (\bibinfo {year}
  {2001})}\BibitemShut {NoStop}%
\bibitem [{\citenamefont {Cheng}\ \emph {et~al.}(2007)\citenamefont {Cheng},
  \citenamefont {C{\'e}pas}, \citenamefont {Leung},\ and\ \citenamefont
  {Ziman}}]{Cheng2007}%
  \BibitemOpen
  \bibfield  {author} {\bibinfo {author} {\bibfnamefont {Y.~F.}\ \bibnamefont
  {Cheng}}, \bibinfo {author} {\bibfnamefont {O.}~\bibnamefont {C{\'e}pas}},
  \bibinfo {author} {\bibfnamefont {P.~W.}\ \bibnamefont {Leung}}, \ and\
  \bibinfo {author} {\bibfnamefont {T.}~\bibnamefont {Ziman}},\ }\href
  {\doibase 10.1103/PhysRevB.75.144422} {\bibfield  {journal} {\bibinfo
  {journal} {Phys. Rev. B}\ }\textbf {\bibinfo {volume} {75}},\ \bibinfo
  {pages} {144422} (\bibinfo {year} {2007})}\BibitemShut {NoStop}%
\bibitem [{\citenamefont {Romh\'anyi}\ \emph {et~al.}(2011)\citenamefont
  {Romh\'anyi}, \citenamefont {Totsuka},\ and\ \citenamefont
  {Penc}}]{Romhanyi2011}%
  \BibitemOpen
  \bibfield  {author} {\bibinfo {author} {\bibfnamefont {J.}~\bibnamefont
  {Romh\'anyi}}, \bibinfo {author} {\bibfnamefont {K.}~\bibnamefont {Totsuka}},
  \ and\ \bibinfo {author} {\bibfnamefont {K.}~\bibnamefont {Penc}},\ }\href
  {\doibase 10.1103/PhysRevB.83.024413} {\bibfield  {journal} {\bibinfo
  {journal} {Phys. Rev. B}\ }\textbf {\bibinfo {volume} {83}},\ \bibinfo
  {pages} {024413} (\bibinfo {year} {2011})}\BibitemShut {NoStop}%
\bibitem [{\citenamefont {Miyahara}\ \emph {et~al.}(2004)\citenamefont
  {Miyahara}, \citenamefont {Mila}, \citenamefont {Kodama}, \citenamefont
  {Takigawa}, \citenamefont {Horvatic}, \citenamefont {Berthier}, \citenamefont
  {Kageyama},\ and\ \citenamefont {Ueda}}]{Miyahara2008}%
  \BibitemOpen
  \bibfield  {author} {\bibinfo {author} {\bibfnamefont {S.}~\bibnamefont
  {Miyahara}}, \bibinfo {author} {\bibfnamefont {F.}~\bibnamefont {Mila}},
  \bibinfo {author} {\bibfnamefont {K.}~\bibnamefont {Kodama}}, \bibinfo
  {author} {\bibfnamefont {M.}~\bibnamefont {Takigawa}}, \bibinfo {author}
  {\bibfnamefont {M.}~\bibnamefont {Horvatic}}, \bibinfo {author}
  {\bibfnamefont {C.}~\bibnamefont {Berthier}}, \bibinfo {author}
  {\bibfnamefont {H.}~\bibnamefont {Kageyama}}, \ and\ \bibinfo {author}
  {\bibfnamefont {Y.}~\bibnamefont {Ueda}},\ }\href
  {http://stacks.iop.org/0953-8984/16/i=11/a=048} {\bibfield  {journal}
  {\bibinfo  {journal} {Journal of Physics: Condensed Matter}\ }\textbf
  {\bibinfo {volume} {16}},\ \bibinfo {pages} {S911} (\bibinfo {year}
  {2004})}\BibitemShut {NoStop}%
\bibitem [{\citenamefont {Smith}\ and\ \citenamefont
  {Keszler}(1991)}]{Smith1991}%
  \BibitemOpen
  \bibfield  {author} {\bibinfo {author} {\bibfnamefont {R.~W.}\ \bibnamefont
  {Smith}}\ and\ \bibinfo {author} {\bibfnamefont {D.~A.}\ \bibnamefont
  {Keszler}},\ }\href {\doibase http://dx.doi.org/10.1016/0022-4596(91)90316-A}
  {\bibfield  {journal} {\bibinfo  {journal} {Journal of Solid State
  Chemistry}\ }\textbf {\bibinfo {volume} {93}},\ \bibinfo {pages} {430 }
  (\bibinfo {year} {1991})}\BibitemShut {NoStop}%
\bibitem [{\citenamefont {Sparta}\ \emph {et~al.}(2001)\citenamefont {Sparta},
  \citenamefont {Redhammer}, \citenamefont {Roussel}, \citenamefont {Heger},
  \citenamefont {Roth}, \citenamefont {Lemmens}, \citenamefont {Ionescu},
  \citenamefont {Grove}, \citenamefont {G{\"{u}}ntherodt}, \citenamefont
  {H{\"{u}}ning}, \citenamefont {Lueken}, \citenamefont {Kageyama},
  \citenamefont {Onizuka},\ and\ \citenamefont {Ueda}}]{sparta2001}%
  \BibitemOpen
  \bibfield  {author} {\bibinfo {author} {\bibfnamefont {K.}~\bibnamefont
  {Sparta}}, \bibinfo {author} {\bibfnamefont {G.}~\bibnamefont {Redhammer}},
  \bibinfo {author} {\bibfnamefont {P.}~\bibnamefont {Roussel}}, \bibinfo
  {author} {\bibfnamefont {G.}~\bibnamefont {Heger}}, \bibinfo {author}
  {\bibfnamefont {G.}~\bibnamefont {Roth}}, \bibinfo {author} {\bibfnamefont
  {P.}~\bibnamefont {Lemmens}}, \bibinfo {author} {\bibfnamefont
  {A.}~\bibnamefont {Ionescu}}, \bibinfo {author} {\bibfnamefont
  {M.}~\bibnamefont {Grove}}, \bibinfo {author} {\bibfnamefont
  {G.}~\bibnamefont {G{\"{u}}ntherodt}}, \bibinfo {author} {\bibfnamefont
  {F.}~\bibnamefont {H{\"{u}}ning}}, \bibinfo {author} {\bibfnamefont
  {H.}~\bibnamefont {Lueken}}, \bibinfo {author} {\bibfnamefont
  {H.}~\bibnamefont {Kageyama}}, \bibinfo {author} {\bibfnamefont
  {K.}~\bibnamefont {Onizuka}}, \ and\ \bibinfo {author} {\bibfnamefont
  {Y.}~\bibnamefont {Ueda}},\ }\href {\doibase 10.1007/s100510170296}
  {\bibfield  {journal} {\bibinfo  {journal} {Eur. Phys. J. B}\ }\textbf
  {\bibinfo {volume} {19}},\ \bibinfo {pages} {507} (\bibinfo {year}
  {2001})}\BibitemShut {NoStop}%
\bibitem [{\citenamefont {Matsuda}\ \emph {et~al.}(2013)\citenamefont
  {Matsuda}, \citenamefont {Abe}, \citenamefont {Takeyama}, \citenamefont
  {Kageyama}, \citenamefont {Corboz}, \citenamefont {Honecker}, \citenamefont
  {Manmana}, \citenamefont {Foltin}, \citenamefont {Schmidt},\ and\
  \citenamefont {Mila}}]{Matsuda2013}%
  \BibitemOpen
  \bibfield  {author} {\bibinfo {author} {\bibfnamefont {Y.~H.}\ \bibnamefont
  {Matsuda}}, \bibinfo {author} {\bibfnamefont {N.}~\bibnamefont {Abe}},
  \bibinfo {author} {\bibfnamefont {S.}~\bibnamefont {Takeyama}}, \bibinfo
  {author} {\bibfnamefont {H.}~\bibnamefont {Kageyama}}, \bibinfo {author}
  {\bibfnamefont {P.}~\bibnamefont {Corboz}}, \bibinfo {author} {\bibfnamefont
  {A.}~\bibnamefont {Honecker}}, \bibinfo {author} {\bibfnamefont {S.~R.}\
  \bibnamefont {Manmana}}, \bibinfo {author} {\bibfnamefont {G.~R.}\
  \bibnamefont {Foltin}}, \bibinfo {author} {\bibfnamefont {K.~P.}\
  \bibnamefont {Schmidt}}, \ and\ \bibinfo {author} {\bibfnamefont
  {F.}~\bibnamefont {Mila}},\ }\href {\doibase 10.1103/PhysRevLett.111.137204}
  {\bibfield  {journal} {\bibinfo  {journal} {Phys. Rev. Lett.}\ }\textbf
  {\bibinfo {volume} {111}},\ \bibinfo {pages} {137204} (\bibinfo {year}
  {2013})}\BibitemShut {NoStop}%
\bibitem [{\citenamefont {Koga}\ and\ \citenamefont
  {Kawakami}(2000)}]{Koga2000}%
  \BibitemOpen
  \bibfield  {author} {\bibinfo {author} {\bibfnamefont {A.}~\bibnamefont
  {Koga}}\ and\ \bibinfo {author} {\bibfnamefont {N.}~\bibnamefont
  {Kawakami}},\ }\href {\doibase 10.1103/PhysRevLett.84.4461} {\bibfield
  {journal} {\bibinfo  {journal} {Phys. Rev. Lett.}\ }\textbf {\bibinfo
  {volume} {84}},\ \bibinfo {pages} {4461} (\bibinfo {year}
  {2000})}\BibitemShut {NoStop}%
\bibitem [{\citenamefont {Corboz}\ and\ \citenamefont
  {Mila}(2013)}]{Corboz2013}%
  \BibitemOpen
  \bibfield  {author} {\bibinfo {author} {\bibfnamefont {P.}~\bibnamefont
  {Corboz}}\ and\ \bibinfo {author} {\bibfnamefont {F.}~\bibnamefont {Mila}},\
  }\href {\doibase 10.1103/PhysRevB.87.115144} {\bibfield  {journal} {\bibinfo
  {journal} {Phys. Rev. B}\ }\textbf {\bibinfo {volume} {87}},\ \bibinfo
  {pages} {115144} (\bibinfo {year} {2013})}\BibitemShut {NoStop}%
\bibitem [{\citenamefont {Bernevig}\ and\ \citenamefont
  {Hughes}(2013)}]{Bernevig}%
  \BibitemOpen
  \bibfield  {author} {\bibinfo {author} {\bibfnamefont {B.~A.}\ \bibnamefont
  {Bernevig}}\ and\ \bibinfo {author} {\bibfnamefont {T.~L.}\ \bibnamefont
  {Hughes}},\ }\enquote {\bibinfo {title} {Topological insulators and
  topological superconductors},}\ \ (\bibinfo  {publisher} {Princeton
  University Press},\ \bibinfo {year} {2013})\ Chap.~\bibinfo {chapter} {8},
  p.~\bibinfo {pages} {96}\BibitemShut {NoStop}%
\bibitem [{\citenamefont {Hatsugai}(1993)}]{Hatsugai1993}%
  \BibitemOpen
  \bibfield  {author} {\bibinfo {author} {\bibfnamefont {Y.}~\bibnamefont
  {Hatsugai}},\ }\href {\doibase 10.1103/PhysRevLett.71.3697} {\bibfield
  {journal} {\bibinfo  {journal} {Phys. Rev. Lett.}\ }\textbf {\bibinfo
  {volume} {71}},\ \bibinfo {pages} {3697} (\bibinfo {year}
  {1993})}\BibitemShut {NoStop}%
\bibitem [{\citenamefont {Sundaram}\ and\ \citenamefont
  {Niu}(1999)}]{Sundaram1999}%
  \BibitemOpen
  \bibfield  {author} {\bibinfo {author} {\bibfnamefont {G.}~\bibnamefont
  {Sundaram}}\ and\ \bibinfo {author} {\bibfnamefont {Q.}~\bibnamefont {Niu}},\
  }\href {\doibase 10.1103/PhysRevB.59.14915} {\bibfield  {journal} {\bibinfo
  {journal} {Phys. Rev. B}\ }\textbf {\bibinfo {volume} {59}},\ \bibinfo
  {pages} {14915} (\bibinfo {year} {1999})}\BibitemShut {NoStop}%
\bibitem [{\citenamefont {Xiao}\ \emph {et~al.}(2010)\citenamefont {Xiao},
  \citenamefont {Chang},\ and\ \citenamefont {Niu}}]{Niu2010}%
  \BibitemOpen
  \bibfield  {author} {\bibinfo {author} {\bibfnamefont {D.}~\bibnamefont
  {Xiao}}, \bibinfo {author} {\bibfnamefont {M.-C.}\ \bibnamefont {Chang}}, \
  and\ \bibinfo {author} {\bibfnamefont {Q.}~\bibnamefont {Niu}},\ }\href
  {\doibase 10.1103/RevModPhys.82.1959} {\bibfield  {journal} {\bibinfo
  {journal} {Rev. Mod. Phys.}\ }\textbf {\bibinfo {volume} {82}},\ \bibinfo
  {pages} {1959} (\bibinfo {year} {2010})}\BibitemShut {NoStop}%
\bibitem [{\citenamefont {Apaja}\ \emph {et~al.}(2010)\citenamefont {Apaja},
  \citenamefont {Hyrk\"as},\ and\ \citenamefont {Manninen}}]{Apaja2010}%
  \BibitemOpen
  \bibfield  {author} {\bibinfo {author} {\bibfnamefont {V.}~\bibnamefont
  {Apaja}}, \bibinfo {author} {\bibfnamefont {M.}~\bibnamefont {Hyrk\"as}}, \
  and\ \bibinfo {author} {\bibfnamefont {M.}~\bibnamefont {Manninen}},\ }\href
  {\doibase 10.1103/PhysRevA.82.041402} {\bibfield  {journal} {\bibinfo
  {journal} {Phys. Rev. A}\ }\textbf {\bibinfo {volume} {82}},\ \bibinfo
  {pages} {041402} (\bibinfo {year} {2010})}\BibitemShut {NoStop}%
\bibitem [{\citenamefont {Huang}\ \emph {et~al.}(2011)\citenamefont {Huang},
  \citenamefont {Lai}, \citenamefont {Zheng},\ and\ \citenamefont
  {Chan}}]{Huang2011}%
  \BibitemOpen
  \bibfield  {author} {\bibinfo {author} {\bibfnamefont {X.}~\bibnamefont
  {Huang}}, \bibinfo {author} {\bibfnamefont {Y.}~\bibnamefont {Lai}}, \bibinfo
  {author} {\bibfnamefont {H.}~\bibnamefont {Zheng}}, \ and\ \bibinfo {author}
  {\bibfnamefont {C.~T.}\ \bibnamefont {Chan}},\ }\href {\doibase
  doi:10.1038/nmat3030} {\bibfield  {journal} {\bibinfo  {journal} {Nature
  Materials}\ }\textbf {\bibinfo {volume} {10}},\ \bibinfo {pages} {582}
  (\bibinfo {year} {2011})}\BibitemShut {NoStop}%
\bibitem [{\citenamefont {Asano}\ and\ \citenamefont
  {Hotta}(2011)}]{Asano2011}%
  \BibitemOpen
  \bibfield  {author} {\bibinfo {author} {\bibfnamefont {K.}~\bibnamefont
  {Asano}}\ and\ \bibinfo {author} {\bibfnamefont {C.}~\bibnamefont {Hotta}},\
  }\href {\doibase 10.1103/PhysRevB.83.245125} {\bibfield  {journal} {\bibinfo
  {journal} {Phys. Rev. B}\ }\textbf {\bibinfo {volume} {83}},\ \bibinfo
  {pages} {245125} (\bibinfo {year} {2011})}\BibitemShut {NoStop}%
\bibitem [{\citenamefont {D\'ora}\ \emph {et~al.}(2011)\citenamefont {D\'ora},
  \citenamefont {Kailasvuori},\ and\ \citenamefont {Moessner}}]{Dora2011}%
  \BibitemOpen
  \bibfield  {author} {\bibinfo {author} {\bibfnamefont {B.}~\bibnamefont
  {D\'ora}}, \bibinfo {author} {\bibfnamefont {J.}~\bibnamefont {Kailasvuori}},
  \ and\ \bibinfo {author} {\bibfnamefont {R.}~\bibnamefont {Moessner}},\
  }\href {\doibase 10.1103/PhysRevB.84.195422} {\bibfield  {journal} {\bibinfo
  {journal} {Phys. Rev. B}\ }\textbf {\bibinfo {volume} {84}},\ \bibinfo
  {pages} {195422} (\bibinfo {year} {2011})}\BibitemShut {NoStop}%
\bibitem [{\citenamefont {Yamashita}\ \emph {et~al.}(2013)\citenamefont
  {Yamashita}, \citenamefont {Tomura}, \citenamefont {Yanagi},\ and\
  \citenamefont {Ueda}}]{Yamashita2013}%
  \BibitemOpen
  \bibfield  {author} {\bibinfo {author} {\bibfnamefont {Y.}~\bibnamefont
  {Yamashita}}, \bibinfo {author} {\bibfnamefont {M.}~\bibnamefont {Tomura}},
  \bibinfo {author} {\bibfnamefont {Y.}~\bibnamefont {Yanagi}}, \ and\ \bibinfo
  {author} {\bibfnamefont {K.}~\bibnamefont {Ueda}},\ }\href {\doibase
  10.1103/PhysRevB.88.195104} {\bibfield  {journal} {\bibinfo  {journal} {Phys.
  Rev. B}\ }\textbf {\bibinfo {volume} {88}},\ \bibinfo {pages} {195104}
  (\bibinfo {year} {2013})}\BibitemShut {NoStop}%
\bibitem [{\citenamefont {Matan}\ \emph {et~al.}(2010)\citenamefont {Matan},
  \citenamefont {Ono}, \citenamefont {Fukumoto}, \citenamefont {Sato},
  \citenamefont {Yamaura}, \citenamefont {Yano}, \citenamefont {Morita},\ and\
  \citenamefont {Tanaka}}]{Matan2010}%
  \BibitemOpen
  \bibfield  {author} {\bibinfo {author} {\bibfnamefont {K.}~\bibnamefont
  {Matan}}, \bibinfo {author} {\bibfnamefont {T.}~\bibnamefont {Ono}}, \bibinfo
  {author} {\bibfnamefont {Y.}~\bibnamefont {Fukumoto}}, \bibinfo {author}
  {\bibfnamefont {T.~J.}\ \bibnamefont {Sato}}, \bibinfo {author}
  {\bibfnamefont {J.}~\bibnamefont {Yamaura}}, \bibinfo {author} {\bibfnamefont
  {M.}~\bibnamefont {Yano}}, \bibinfo {author} {\bibfnamefont {K.}~\bibnamefont
  {Morita}}, \ and\ \bibinfo {author} {\bibfnamefont {H.}~\bibnamefont
  {Tanaka}},\ }\href {\doibase doi:10.1038/nphys1761} {\bibfield  {journal}
  {\bibinfo  {journal} {Nature Physics}\ }\textbf {\bibinfo {volume} {6}},\
  \bibinfo {pages} {865} (\bibinfo {year} {2010})}\BibitemShut {NoStop}%
\bibitem [{\citenamefont {Hwang}\ \emph {et~al.}(2012)\citenamefont {Hwang},
  \citenamefont {Park},\ and\ \citenamefont {Kim}}]{Kyusung2012}%
  \BibitemOpen
  \bibfield  {author} {\bibinfo {author} {\bibfnamefont {K.}~\bibnamefont
  {Hwang}}, \bibinfo {author} {\bibfnamefont {K.}~\bibnamefont {Park}}, \ and\
  \bibinfo {author} {\bibfnamefont {Y.~B.}\ \bibnamefont {Kim}},\ }\href
  {\doibase 10.1103/PhysRevB.86.214407} {\bibfield  {journal} {\bibinfo
  {journal} {Phys. Rev. B}\ }\textbf {\bibinfo {volume} {86}},\ \bibinfo
  {pages} {214407} (\bibinfo {year} {2012})}\BibitemShut {NoStop}%
\bibitem [{\citenamefont {Tovar}\ \emph {et~al.}(2009)\citenamefont {Tovar},
  \citenamefont {Raman},\ and\ \citenamefont {Shtengel}}]{Tover2009}%
  \BibitemOpen
  \bibfield  {author} {\bibinfo {author} {\bibfnamefont {M.}~\bibnamefont
  {Tovar}}, \bibinfo {author} {\bibfnamefont {K.~S.}\ \bibnamefont {Raman}}, \
  and\ \bibinfo {author} {\bibfnamefont {K.}~\bibnamefont {Shtengel}},\ }\href
  {\doibase 10.1103/PhysRevB.79.024405} {\bibfield  {journal} {\bibinfo
  {journal} {Phys. Rev. B}\ }\textbf {\bibinfo {volume} {79}},\ \bibinfo
  {pages} {024405} (\bibinfo {year} {2009})}\BibitemShut {NoStop}%
\end{thebibliography}

\begin{thebibliography}{8}%
\makeatletter
\providecommand \@ifxundefined [1]{%
 \@ifx{#1\undefined}
}%
\providecommand \@ifnum [1]{%
 \ifnum #1\expandafter \@firstoftwo
 \else \expandafter \@secondoftwo
 \fi
}%
\providecommand \@ifx [1]{%
 \ifx #1\expandafter \@firstoftwo
 \else \expandafter \@secondoftwo
 \fi
}%
\providecommand \natexlab [1]{#1}%
\providecommand \enquote  [1]{``#1''}%
\providecommand \bibnamefont  [1]{#1}%
\providecommand \bibfnamefont [1]{#1}%
\providecommand \citenamefont [1]{#1}%
\providecommand \href@noop [0]{\@secondoftwo}%
\providecommand \href [0]{\begingroup \@sanitize@url \@href}%
\providecommand \@href[1]{\@@startlink{#1}\@@href}%
\providecommand \@@href[1]{\endgroup#1\@@endlink}%
\providecommand \@sanitize@url [0]{\catcode `\\12\catcode `\$12\catcode
  `\&12\catcode `\#12\catcode `\^12\catcode `\_12\catcode `\%12\relax}%
\providecommand \@@startlink[1]{}%
\providecommand \@@endlink[0]{}%
\providecommand \url  [0]{\begingroup\@sanitize@url \@url }%
\providecommand \@url [1]{\endgroup\@href {#1}{\urlprefix }}%
\providecommand \urlprefix  [0]{URL }%
\providecommand \Eprint [0]{\href }%
\providecommand \doibase [0]{http://dx.doi.org/}%
\providecommand \selectlanguage [0]{\@gobble}%
\providecommand \bibinfo  [0]{\@secondoftwo}%
\providecommand \bibfield  [0]{\@secondoftwo}%
\providecommand \translation [1]{[#1]}%
\providecommand \BibitemOpen [0]{}%
\providecommand \bibitemStop [0]{}%
\providecommand \bibitemNoStop [0]{.\EOS\space}%
\providecommand \EOS [0]{\spacefactor3000\relax}%
\providecommand \BibitemShut  [1]{\csname bibitem#1\endcsname}%
\let\auto@bib@innerbib\@empty
\bibitem [{\citenamefont {Choi}\ \emph {et~al.}(2003)\citenamefont {Choi},
  \citenamefont {Pashkevich}, \citenamefont {Lamonova}, \citenamefont
  {Kageyama}, \citenamefont {Ueda},\ and\ \citenamefont {Lemmens}}]{Choi2003}%
  \BibitemOpen
  \bibfield  {author} {\bibinfo {author} {\bibfnamefont {K.-Y.}\ \bibnamefont
  {Choi}}, \bibinfo {author} {\bibfnamefont {Y.~G.}\ \bibnamefont
  {Pashkevich}}, \bibinfo {author} {\bibfnamefont {K.~V.}\ \bibnamefont
  {Lamonova}}, \bibinfo {author} {\bibfnamefont {H.}~\bibnamefont {Kageyama}},
  \bibinfo {author} {\bibfnamefont {Y.}~\bibnamefont {Ueda}}, \ and\ \bibinfo
  {author} {\bibfnamefont {P.}~\bibnamefont {Lemmens}},\ }\href {\doibase
  10.1103/PhysRevB.68.104418} {\bibfield  {journal} {\bibinfo  {journal} {Phys.
  Rev. B}\ }\textbf {\bibinfo {volume} {68}},\ \bibinfo {pages} {104418}
  (\bibinfo {year} {2003})}\BibitemShut {NoStop}%
\bibitem [{\citenamefont {Kodama}\ \emph {et~al.}(2005)\citenamefont {Kodama},
  \citenamefont {Miyahara}, \citenamefont {Takigawa}, \citenamefont
  {Horvatić}, \citenamefont {Berthier}, \citenamefont {Mila}, \citenamefont
  {Kageyama},\ and\ \citenamefont {Ueda}}]{Kodama2005}%
  \BibitemOpen
  \bibfield  {author} {\bibinfo {author} {\bibfnamefont {K.}~\bibnamefont
  {Kodama}}, \bibinfo {author} {\bibfnamefont {S.}~\bibnamefont {Miyahara}},
  \bibinfo {author} {\bibfnamefont {M.}~\bibnamefont {Takigawa}}, \bibinfo
  {author} {\bibfnamefont {M.}~\bibnamefont {Horvatić}}, \bibinfo {author}
  {\bibfnamefont {C.}~\bibnamefont {Berthier}}, \bibinfo {author}
  {\bibfnamefont {F.}~\bibnamefont {Mila}}, \bibinfo {author} {\bibfnamefont
  {H.}~\bibnamefont {Kageyama}}, \ and\ \bibinfo {author} {\bibfnamefont
  {Y.}~\bibnamefont {Ueda}},\ }\href
  {http://stacks.iop.org/0953-8984/17/i=4/a=L02} {\bibfield  {journal}
  {\bibinfo  {journal} {Journal of Physics: Condensed Matter}\ }\textbf
  {\bibinfo {volume} {17}},\ \bibinfo {pages} {L61} (\bibinfo {year}
  {2005})}\BibitemShut {NoStop}%
\bibitem [{\citenamefont {Cheng}\ \emph {et~al.}()\citenamefont {Cheng},
  \citenamefont {C{\'e}pas}, \citenamefont {Leung},\ and\ \citenamefont
  {Ziman}}]{Cheng2007}%
  \BibitemOpen
  \bibfield  {author} {\bibinfo {author} {\bibfnamefont {Y.~F.}\ \bibnamefont
  {Cheng}}, \bibinfo {author} {\bibfnamefont {O.}~\bibnamefont {C{\'e}pas}},
  \bibinfo {author} {\bibfnamefont {P.~W.}\ \bibnamefont {Leung}}, \ and\
  \bibinfo {author} {\bibfnamefont {T.}~\bibnamefont {Ziman}},\ }\href
  {\doibase 10.1103/PhysRevB.75.144422} {\bibinfo  {journal} {Phys. Rev. B}\ ,\
  \bibinfo {pages} {144422}}\BibitemShut {NoStop}%
\bibitem [{\citenamefont {Bernevig}\ and\ \citenamefont
  {Hughes}(2013)}]{Bernevig}%
  \BibitemOpen
\bibfield  {journal} {  }\bibfield  {author} {\bibinfo {author} {\bibfnamefont
  {B.~A.}\ \bibnamefont {Bernevig}}\ and\ \bibinfo {author} {\bibfnamefont
  {T.~L.}\ \bibnamefont {Hughes}},\ }\href@noop {} {\emph {\bibinfo {title}
  {Topological Insulators and Topological Superconductors}}}\ (\bibinfo
  {publisher} {Princeton University Press},\ \bibinfo {year}
  {2013})\BibitemShut {NoStop}%
\bibitem [{\citenamefont {Zayed}\ \emph {et~al.}(2014)\citenamefont {Zayed},
  \citenamefont {R\"uegg}, \citenamefont {Str\"assle}, \citenamefont {Stuhr},
  \citenamefont {Roessli}, \citenamefont {Ay}, \citenamefont {Mesot},
  \citenamefont {Link}, \citenamefont {Pomjakushina}, \citenamefont
  {Stingaciu}, \citenamefont {Conder},\ and\ \citenamefont
  {R\o{}nnow}}]{Ronnow2014}%
  \BibitemOpen
  \bibfield  {author} {\bibinfo {author} {\bibfnamefont {M.~E.}\ \bibnamefont
  {Zayed}}, \bibinfo {author} {\bibfnamefont {C.}~\bibnamefont {R\"uegg}},
  \bibinfo {author} {\bibfnamefont {T.}~\bibnamefont {Str\"assle}}, \bibinfo
  {author} {\bibfnamefont {U.}~\bibnamefont {Stuhr}}, \bibinfo {author}
  {\bibfnamefont {B.}~\bibnamefont {Roessli}}, \bibinfo {author} {\bibfnamefont
  {M.}~\bibnamefont {Ay}}, \bibinfo {author} {\bibfnamefont {J.}~\bibnamefont
  {Mesot}}, \bibinfo {author} {\bibfnamefont {P.}~\bibnamefont {Link}},
  \bibinfo {author} {\bibfnamefont {E.}~\bibnamefont {Pomjakushina}}, \bibinfo
  {author} {\bibfnamefont {M.}~\bibnamefont {Stingaciu}}, \bibinfo {author}
  {\bibfnamefont {K.}~\bibnamefont {Conder}}, \ and\ \bibinfo {author}
  {\bibfnamefont {H.~M.}\ \bibnamefont {R\o{}nnow}},\ }\href {\doibase
  10.1103/PhysRevLett.113.067201} {\bibfield  {journal} {\bibinfo  {journal}
  {Phys. Rev. Lett.}\ }\textbf {\bibinfo {volume} {113}},\ \bibinfo {pages}
  {067201} (\bibinfo {year} {2014})}\BibitemShut {NoStop}%
\bibitem [{\citenamefont {Gaulin}\ \emph {et~al.}()\citenamefont {Gaulin},
  \citenamefont {Lee}, \citenamefont {Haravifard}, \citenamefont {Castellan},
  \citenamefont {Berlinsky}, \citenamefont {Dabkowska}, \citenamefont {Qiu},\
  and\ \citenamefont {Copley}}]{Gaulin2004}%
  \BibitemOpen
  \bibfield  {author} {\bibinfo {author} {\bibfnamefont {B.~D.}\ \bibnamefont
  {Gaulin}}, \bibinfo {author} {\bibfnamefont {S.~H.}\ \bibnamefont {Lee}},
  \bibinfo {author} {\bibfnamefont {S.}~\bibnamefont {Haravifard}}, \bibinfo
  {author} {\bibfnamefont {J.~P.}\ \bibnamefont {Castellan}}, \bibinfo {author}
  {\bibfnamefont {A.~J.}\ \bibnamefont {Berlinsky}}, \bibinfo {author}
  {\bibfnamefont {H.~A.}\ \bibnamefont {Dabkowska}}, \bibinfo {author}
  {\bibfnamefont {Y.}~\bibnamefont {Qiu}}, \ and\ \bibinfo {author}
  {\bibfnamefont {J.~R.~D.}\ \bibnamefont {Copley}},\ }\href {\doibase
  10.1103/PhysRevLett.93.267202} {\bibinfo  {journal} {Phys. Rev. Lett.}\ ,\
  \bibinfo {pages} {267202}}\BibitemShut {NoStop}%
\bibitem [{\citenamefont {Miyahara}\ and\ \citenamefont
  {Ueda}(1999)}]{Miyahara1999}%
  \BibitemOpen
\bibfield  {journal} {  }\bibfield  {author} {\bibinfo {author} {\bibfnamefont
  {S.}~\bibnamefont {Miyahara}}\ and\ \bibinfo {author} {\bibfnamefont
  {K.}~\bibnamefont {Ueda}},\ }\href {\doibase 10.1103/PhysRevLett.82.3701}
  {\bibfield  {journal} {\bibinfo  {journal} {Phys. Rev. Lett.}\ }\textbf
  {\bibinfo {volume} {82}},\ \bibinfo {pages} {3701} (\bibinfo {year}
  {1999})}\BibitemShut {NoStop}%
\bibitem [{\citenamefont {Weihong}\ \emph {et~al.}(1999)\citenamefont
  {Weihong}, \citenamefont {Hamer},\ and\ \citenamefont
  {Oitmaa}}]{Weihong1999}%
  \BibitemOpen
  \bibfield  {author} {\bibinfo {author} {\bibfnamefont {Z.}~\bibnamefont
  {Weihong}}, \bibinfo {author} {\bibfnamefont {C.~J.}\ \bibnamefont {Hamer}},
  \ and\ \bibinfo {author} {\bibfnamefont {J.}~\bibnamefont {Oitmaa}},\ }\href
  {\doibase 10.1103/PhysRevB.60.6608} {\bibfield  {journal} {\bibinfo
  {journal} {Phys. Rev. B}\ }\textbf {\bibinfo {volume} {60}},\ \bibinfo
  {pages} {6608} (\bibinfo {year} {1999})}\BibitemShut {NoStop}%
\end{thebibliography}
%

\end{document}